\documentclass{article}
\usepackage{natbib}

\usepackage[utf8]{inputenc} 
\usepackage[T1]{fontenc}    
\usepackage[dvipsnames]{xcolor}
\definecolor{SlateGrey}{HTML}{2F4F4F}\usepackage[colorlinks=true,citecolor=SlateGrey,linkcolor=blue,urlcolor=blue]{hyperref}
\usepackage{url}            
\usepackage{booktabs}       
\usepackage{amsfonts}       
\usepackage{nicefrac}       
\usepackage{microtype}      
\usepackage{lipsum}		

\usepackage{doi}

\usepackage{amsmath}
\usepackage{amsfonts}
\usepackage{dsfont}
\usepackage{graphicx}
\usepackage{algorithm}
\usepackage{comment}
\usepackage{arydshln}
\usepackage[title]{appendix}
\usepackage{algpseudocode}
\usepackage{wrapfig}
\setcitestyle{square}
\usepackage{amsthm}
\usepackage{enumitem}
\usepackage{subfig}
\usepackage{mathtools}
\usepackage{authblk}
\usepackage{filecontents}

\definecolor{mygreen}{rgb}{0.13,0.55,0.13}

\usepackage{blindtext}
\usepackage{geometry}
\title{\LARGE	Data Assimilation with Machine Learning  Surrogate Models: \\  A Case Study with FourCastNet}

\author[1]{Melissa Adrian\thanks{Corresponding author: \url{maadrian@uchicago.edu}}}
\author[1]{Daniel Sanz-Alonso}
\author[1,2]{Rebecca Willett}

\affil[1]{Department of Statistics, The University of Chicago, Chicago, IL 60637, USA}
\affil[2]{Department of Computer Science, The University of Chicago, Chicago, IL 60637, USA}

\date{}

\begin{document}
\maketitle

\begin{abstract}
Modern data-driven surrogate models for weather forecasting  provide accurate short-term predictions but inaccurate and nonphysical long-term forecasts. This paper investigates online weather prediction using machine learning surrogates supplemented with partial and noisy observations. We empirically demonstrate and theoretically justify that, despite the long-time instability of the surrogates and the sparsity of the observations, filtering estimates can remain accurate in the long-time horizon. As a case study, we integrate FourCastNet, a weather surrogate model, within a variational data assimilation framework using partial, noisy ERA5 data. Our results show that filtering estimates remain accurate over a year-long assimilation window and provide effective initial conditions for forecasting tasks, including extreme event prediction.  

\end{abstract}

\section{Introduction}\label{sec:introduction}

Numerical weather prediction (NWP) at an operational scale relies on large-scale systems of partial differential equations to model atmospheric dynamics. However, these physics-based models  
are computationally expensive to simulate, particularly when operating at a high resolution. This computational burden plagues both weather and climate models alike \citep{Tollefson2023}, leading research focus to shift towards cheaper alternatives: 
data-driven machine learning surrogate models for weather forecasting. 

Weather surrogate models have been analyzed and evaluated extensively using high-resolution accurate datasets, usually ERA5 reanalysis data \citep{Hersbach2020}. 
However, in practical weather forecasting settings, we must provide high-fidelity forecasts given only sparse  
observations, often contaminated with measurement errors.
Because major advances in high-resolution data-driven global weather modeling have only been made in the past few years, substantial work has yet to be done to analyze its utility in settings of operational interest, including both data assimilation and forecasting with these sparse and noisy observations.

Data assimilation is an operational task with a long history in weather forecasting, 
rooted in seminal works, such as  \citet{Richardson1922},   \citet{Kalman1960}, and \citet{Gandin1966}. Preliminary data assimilation methods were developed specifically for the vast problem that characterizes weather settings: inferring a high-resolution representation of the atmosphere with only (1) sparse, noisy measurements throughout the globe and (2) a time evolution mapping of atmospheric states. Data assimilation produces high-dimensional representations, also referred to as analyses, which provide a detailed view of historical global weather patterns. These analyses can be used for numerous purposes, most notably to initialize forecasts based on current and historical observations. However, producing analyses using physics-based NWP models is computationally expensive, especially for long time horizons. Consequently, cheap-to-evaluate surrogate weather models have enormous potential to expedite this data assimilation process.

 The potential of weather surrogate models to accelerate extreme event prediction has recently received attention at the government level, with an April 2024 U.S. Executive Office report calling attention to its potential widespread operational use \citep{WhiteHouse2024}. This report stated that in the U.S. in 2023 alone, the economic damage due to extreme weather events totalled to \$92.9 billion from 28 weather disasters, and the frequency of these extreme events is expected to continue to increase in the coming years \citep{WhiteHouse2024}. This staggering economic loss and the projected increased frequency in weather disasters point to the escalating need for real-time, accurate weather forecasting. 

Surrogate weather models allow for real-time forecasting and uncertainty quantification given an accurate estimate of the initial or current weather state, which could be especially impactful for time-sensitive extreme events like hurricanes, typhoons, cyclones, etc. In the event of extreme weather, it is paramount to produce, in real time, accurate estimates  of the state of the weather to initialize forecasts, which can be accomplished via data assimilation. Moreover, quickly computing large ensembles of forecasts using these weather surrogates allows for a timely characterization of the distribution of potential future outcomes. Speed, accuracy, and uncertainty quantification are crucial to gauge the severity and likelihood of these potential outcomes and subsequently inform timely decisions regarding public safety.

In our work, we aim to assess the utility of weather surrogate models, particularly a weather surrogate model, FourCastNet \citep{Pathak2022}, in the three main tasks of operational interest: (1) estimation of the high-dimensional weather state from low-resolution observations via a variational data assimilation framework,
(2) multi-step-ahead forecasting using these estimates, and as a more focused forecasting task, (3) extreme event prediction using these estimates. Our high-dimensional state consists of 20 atmospheric features at various pressure levels in the atmosphere at a global $0.25^\circ$ resolution, corresponding to a $720\times1440$ grid. We utilize (1) FourCastNet as our weather model, (2) low-resolution, noisy data derived from the ERA5 reanalysis \citep{Hersbach2020} as a proxy for real observations, and (3) a 3DVar variational data assimilation algorithm \citep{Lorenc2000} to combine forecasts and observations. We demonstrate that pairing FourCastNet within 3DVar to assimilate low-resolution, noisy observational data can produce visually realistic weather patterns while maintaining a stable reconstruction error over a long time horizon, and we provide a theoretical justification for this result. Additionally, we show that these 3DVar analyses can serve as effective initial conditions for forecasting tasks, even in our extreme event case study.

\section{Contributions}\label{sec:contributions} 
This work provides new empirical and theoretical evidence that data-driven surrogate models for weather forecasting can be successful in data assimilation tasks. We illustrate the potential of purely data-driven, global weather data assimilation using a current weather surrogate model, FourCastNet, and low-resolution, noisy data. Our assimilation is based on a simple 3DVar filter that can be run with a single NVIDIA A100 GPU. The main contributions of this paper are to:
\begin{enumerate}
     \item Empirically show the accuracy of 3DVar filtering with FourCastNet over a year-long assimilation window given a sufficiently rich set of low-resolution, noisy observations. 
     \item Rigorously prove long-time assimilation 
    accuracy of 3DVar with a short-time accurate surrogate forecast model and a sufficiently rich set of partial, noisy observations.
    \item Demonstrate that filtering estimates provide successful initial conditions for forecasting tasks, including extreme event prediction. 

\end{enumerate}

\section{Related work}\label{sec:related_work}

In recent years, there have been substantial advances in machine learning to create surrogates of numerical weather models in order to produce predictions at a fraction of the cost while maintaining short-term accuracy. Notable global data-driven weather models include FourCastNet versions 1 and 2 \citep{Pathak2022, Bonev2023sfno}, 
Pangu-Weather \citep{Bi2023}, GraphCast \citep{Lam2023}, a graph-based weather model from \citep{keisler2022}, FengWu \citep{chen2023fengwu}, and FuXi \citep{chen2023fuxi}. In this work, we focus on evaluating FourCastNet as a case study.

In addition to training high-fidelity data-driven weather models, machine learning for weather and climate applications is rapidly advancing in many other directions. \citet{Krasnopolsky2023} provides a current overview of advances in machine learning for data assimilation, modeling physics, and post-processing for weather and climate systems. Additionally, \citet{Cheng2023}, \citet[Chapter 10]{Chen2023}, and \citet{Bocquet2023} survey current approaches to combine machine learning with data assimilation, including the use of surrogate forecast models that is the focus of our work.

\paragraph{Data-driven weather forecasting in data assimilation} Some recent works have explored replacing numerical weather prediction models with data-driven prediction models trained on smaller-scale datasets. For instance, \citet{Chattopadhyay2022} develops a data-driven weather prediction architecture for  a 5.625$^\circ$ resolution representation of global geopotential at 500 hPa, corresponding to a 32$\times 64$ latitude/longitude grid. This work shows success in providing analyses that accurately estimate the ground truth using a sigma-point ensemble Kalman filter \citep{Tang2014} paired with their novel architecture. Another example is \citet{Maulik2022}, which builds a surrogate model of geopotential height at 500 hPa over North America. 
\citet{Maulik2022} shows that using their trained surrogate model within a 4DVar algorithm \citep{Dimet1986} produces analyses that, when used as initial conditions for forecasting tasks, outperform forecasts that are initialized based on climatology.

The successes on smaller scale data point to potential success for larger scale integration of weather surrogates in data assimilation tasks. 
\citet{Huang2024} and \citet{Xiao2023} are among the first works to combine data assimilation and a weather surrogate at a high dimensional scale. 
\citet{Huang2024} performs a data assimilation task at an operational scale by reconstructing 24 atmospheric features at a $0.25^\circ$ resolution (721$\times$1440 grid) with the data-driven weather surrogate GraphCast \citep{Lam2023}. In addition, the authors formulate a novel alternative to traditional data assimilation tasks through the use of a learnable diffusion model. \citet{Huang2024}, however, focuses on short assimilation horizons and one-step-ahead forecasting in its experiments and utilizes a diffusion-based architecture, which may suffer from inefficiencies in sample generation speed. Addressing this generation speed issue is an area of active research \citep{chen2024overview}. In our work, we expand upon evaluation tasks to assess the quality of longer forecasts using data assimilation analyses in a computationally efficient manner.

\citet{Xiao2023} combines FenguWu \citep{chen2023fengwu} within a 4DVar assimilation scheme to estimate the state of the atmosphere for 69 atmospheric features at a roughly $1.41^\circ$ spatial resolution, corresponding to a $128\times256$ grid. This work assimilates observations that were created from ERA5 data with randomly applied masks, leading to 15\% of locations being observed \citep{Xiao2023}. Similarly to the results we present in this paper, \citet{Xiao2023} shows stable analysis errors across the span of a year. Our work demonstrates similar error stability for a higher dimensional representation of global weather patterns (0.25$^\circ$ resolution) using lower resolution observations (observing at most 1.5\% of locations) compared to \citet{Xiao2023} for observations acquired at fixed spatial locations.

\paragraph{Extreme event forecasting using data-driven weather surrogates.} As numerous cheap-to-evaluate machine learning weather surrogates have been developed and released to the public in recent years, preliminary research efforts have begun to assess the forecast quality in predicting extreme events, including hurricane, cyclone, and typhoon tracking. To highlight a few of these works and their main takeaways regarding extreme storm forecasting, \citet{Magnusson2023} notes that Pangu-Weather's forecasts of cyclone Eunice in 2022 were lacking small-scale features in the atmospheric fields, under-predicted the maximum wind speed of the cyclone, and showed a faster evolution of the cyclone compared to the NWP forecasts. \citet{Lopez-Gomez} notes that training their data-driven weather surrogate with a mean-squared-error loss, which is commonly used to train weather surrogate models, produced less skillful forecast of extremes and overly-smoothed forecasts, especially for larger lead times. Lastly,  \citet{Charlton-Perez2024} notes that the weather surrogates FourCastNet, FourCastNet v2, Pangu-Weather, and GraphCast failed to produce small scale bands of strong winds in their Storm Ciarán case study and consistently under-predicted wind speed; however, the models showed comparable skill to NWP forecasts up to a 48 hour lead time. 

Two key takeaways have emerged from this expanding body of literature, evidenced by the aforementioned work and others: machine learning weather surrogates generally produce overly smoothed predictions, and these models generally under-predict the intensity of extreme events. This issue is actively being investigated, and diffusion-based generative models like GenCast \citep{Gencast} have shown promise in increasing sharpness and predicting closer to the true extremity of rare events, though at a substantial computational cost. In our evaluation of a particular extreme event, we especially inspect the predicted versus actual severity, as well as the level of detail in the predictions. As a case study, we compute forecasts of Typhoon Mawar in 2023 and assess the U- and V-component wind speed at 10 meters (10m) above Earth's surface, the mean sea level pressure, and the location of the typhoon's eye to characterize the quality of forecasts given various types of initializations, including the data assimilation analyses we produce.

\paragraph{Stability theory of 3DVar accuracy.} Many works have analyzed long-term stability and accuracy of nonlinear filtering algorithms \citep{crisan2011oxford} and  data assimilation techniques \citep{kalnay2003atmospheric} that employ the true model for the dynamics. In particular, a large body of work exemplified by  \citet{hayden2011discrete,sanz2015long,law2016filter} has established long-time filter accuracy for a wide class of atmospheric models building on the rich theory of synchronization in chaotic dynamical systems \citep{pecora1990synchronization}. The key idea is that, while for chaotic systems small errors in state estimation are typically exponentially amplified by the dynamics, this growth of errors can be tamed if sufficiently rich observations of the state are assimilated in an online fashion. 
 
Our main theoretical result, Theorem \ref{thm:stability}, establishes long-time accuracy for a 3DVar filtering algorithm that utilizes a surrogate model of the true dynamics. We assume only that (1) the surrogate model is accurate over one assimilation cycle and (2) the observations are sufficiently rich to achieve long-time filter accuracy with a 3DVar algorithm that employs the true dynamics model. Our result is hence similar to \citet{Moodey2013}, which also establishes accuracy under model error, but in contrast to \citet{Moodey2013}  we place no assumptions on the surrogate model other than short-time accuracy, thus making our theory more directly relevant to the context of complex machine learning surrogates for weather forecasting. 

\section{Data description}
\label{sec:data}

\subsection{ECMWF Reanalysis v5 (ERA5)}

ERA5 is a reanalysis dataset that provides hourly atmosphere, land, and ocean feature estimates produced by the variational data assimilation method 4DVar \citep{Rawlins2007} at a resolution of 0.25$^{\circ}$ using observational weather data from 1979 to present day \citep{Hersbach2020}. Data was pulled from the Copernicus Climate Data Store, \citet{ERA5Land} for land features, and \citet{ERA5Pressure} for pressure level features. We retain only a subset of the atmospheric features in ERA5, specifically total column water vapor (TCWV), geopotential at 50, 500, 850, and 1000 hPa, U-component wind speed at 10m from the surface and at 500, 850, and 1000 hPa, V-component wind speed at 10m from the surface and at 500, 850, and 1000 hPa, relative humidity at 500 and 850 hPa, temperature at 2 meters from the surface and at 500 and 850 hPa, surface pressure (sp), and mean sea level pressure (mslp). Consequently, we retained all the features that FourCastNet was trained to predict. Additionally, we standardized the ERA5 dataset using the same global feature means and standard deviations that were used in training FourCastNet.

As described next, for our assimilation and forecasting tasks, the states we attempt to estimate are the high-resolution ERA5 data across 2023 in its native 0.25$^\circ$ resolution, from which we create low-resolution noisy observations. We emphasize that our chosen time range, ERA5 data for 2023, is disjoint from the time range that FourCastNet was trained and tuned on, ERA5 data from 1979 to 2017 \citep{Pathak2022}.

\paragraph{Ground truth states} $\{x_t^{\text{\rm true}}\}_{t=0}^T$. 
The ground truth state $x_t^{\text{true}}\in \mathbb{R}^{d_x}$ at time $t$ represents 20 atmospheric features
at a $0.25^\circ$ resolution, which results in a state dimension $d_x = 20\times 720\times 1440=20,736,000$ at one time point. 
This state is partially observed every 6 hours for the entirety of 2023, so we consider $T= 1460$ time points (4 observations/day$\ \times \ 365 \text{ days}$). Thus, $t=0$ corresponds to January 1, 2023 at 00:00 UTC, and $t=T$ corresponds to December 31, 2023 at 18:00 UTC. Since we do not ever observe the true ground truth state in reality, we will evaluate our results based on a state-of-the-art proxy of atmospheric states, namely the ERA5 dataset \citep{Hersbach2020}.

\paragraph{Observations $y_t$.}
The observations $y_t$ represent data collected at a lower spatial resolution than the true state $x_t^{\text{\rm true}}$. We seek to estimate the high-resolution weather state from these low-resolution observations.
In our experiments, we generate low-resolution $y_t$ from $x_t^{\text{\rm true}}$ to explore the efficacy of the proposed approach for data assimilation using weather surrogate models. Specifically,
we generate $y_t$  using coarse measurements of the 20 ERA5 atmospheric features. 
To generate these observations, we systematically thin out the ERA5 latitude/longitude grid to retain all atmospheric features at every $k$-th coordinate in the latitude and longitude directions, where we vary $k$ in our assimilation experiments in Section \ref{sec:results}\ref{sec:vary_obs}. We additionally add $\mathcal{N}(0,R)$ distributed noise to this ground truth coarsened ERA5 data to model 
 measurement error, where $R=0.0001 I_{d_y}$ and $\mathcal{N}$ refers to a normal distribution. 
For a given $k$, our observations are $y_t\in\mathbb{R}^{d_{y}}$, where $d_{y}=720/k\times 1440/k\times20=20,736,000/k^2$. We give an interpretation of our choices of $k$ in Table \ref{tab:observation_resolutions}. As an illustrative example, Appendix A 
provides a visualization of ground truth ERA5 data in its native $0.25^\circ$ resolution compared to the 4.5$^\circ$ observations for relative humidity at 500 hPa.

 We note that when we discuss observations in a generic data assimilation setting, we describe them as sparse and noisy, while in our experiments, we describe observations as low-resolution and noisy. This difference in language is to emphasize that in generic assimilation settings with observations taken from real sensors, satellites, etc., these observations can be non-uniformly spaced, with the spacing being different for different weather features. In our experimental settings, however, the observations follow a regular, coarsened grid structure.  

 \begin{table*}
   
   \small
   \centering
   \begin{tabular}{ccccc}
   \toprule
   \textbf{$k$} & 
   \textbf{Resolution} &
   \multicolumn{1}{c}{\textbf{\begin{tabular}[c]{@{}c@{}}Latitude/longitude\\ grid size\end{tabular}}} &
   \multicolumn{1}{c}{\textbf{\begin{tabular}[c]{@{}c@{}}Distance between observations\\ along the equator\end{tabular}}} & 
   \textbf{\% of state observed} \\ 
   \midrule
   8 & $2^\circ$& 90$\times$180 & 222 km & 1.56\% \\
   10 & $2.5^\circ$& 72$\times$144 & 278 km & 1.00\% \\
   18 & $4.5^\circ$& 40$\times$80 & 500 km & 0.31\% \\
   20 & $5^\circ$& 36$\times$72 & 556 km & 0.25\% \\
   \bottomrule \\
   \end{tabular}
   \caption{Table describing the observational dataset resolution with corresponding distances between observations along the equator and percentage of the states observed for each dataset. In each of these datasets, we observe all 20 atmospheric features for every $k$-th location of interest in both the latitude and longitude directions. As reference, these   states have a resolution of 0.25$^\circ$, which corresponds to a distance of 28 km between states along the equator and a latitude/longitude grid size of 720$\times$1440.} 
   \label{tab:observation_resolutions}
\end{table*}

\subsection{High-resolution forecasts (HRES) of the European Centre for Medium-Range Weather Forecasts (ECMWF)'s Integrated Forecasting System (IFS)}

In Section \ref{sec:results}\ref{subsec:mawar}, we utilize an additional dataset: archived high-resolution forecasts of ECMWF's IFS, which we refer to as IFS-HRES. For our forecasting task of Typhoon Mawar, we pulled forecasts at 6 hour intervals of IFS-HRES initialized on May 23, 2023 at 00:00 UTC until May 30, 2023 at 12:00 UTC. The native resolution of the IFS-HRES forecasts are 0.1$^\circ$, but in order to match the resolution of the ERA5 data, we pulled these forecasts at 0.25$^\circ$. Since we use these forecasts in our analysis of Typhoon Mawar, we only retained atmospheric variables relevant to this prediction task, which include mean sea level pressure and  U and V component wind speeds at 10m above the surface. 

\subsection{Observational typhoon data from the International Best Track Archive for Climate Stewardship (IBTrACS)}

Section \ref{sec:results}\ref{subsec:mawar} additionally utilizes IBTrAC observational data from the Joint Typhoon Warning Center, which contains key information about tropical cyclones including their locations and intensities over time. We subset this data to focus specifically on Typhoon Mawar between May 23, 2023 00:00 UTC and May 30, 2023 12:00 UTC. 

\section{Methodology}\label{sec:methodology}
\subsection{Setting}\label{sec:setting}
Our goal is to estimate a high-dimensional gridded representation of atmospheric features, denoted $\{x_t^\text{true}\}_{t\geq 1}$, given observations $\{y_t\}_{t\geq0}$, which are derived from the following setting:
\begin{align}\label{eq:true_dynamics_obs}
    x_t^\text{true} &= \mathcal{F}(x_{t-1}^\text{true}), \\ \nonumber
    y_t &=  H x_t^\text{true} + \eta_t, \quad \eta_t \sim \mathcal{N}(0,R),
\end{align}
where $\mathcal{F}$ is the true dynamics governing the evolution of the state,  $H$ is a linear observation operator, and $R$ is a known measurement error covariance matrix.  
We are interested in online estimation, so that at each time $t$ our estimate of the state $x_t^\text{true}$ should only depend on the observations $\{y_0, \ldots, y_t\}$ available at time $t.$ Again, our goal is to reconstruct these high-dimensional states in 6-hourly increments across the year 2023 for $t=1,\dots,T$, where $T=1460$. We next describe the specific problem settings for our numerical results.

 \paragraph{Observation operator $H$.} The observation operator $H \in \{0,1\}^{d_y\times d_x}$ is 
 a subset of the rows of the identity matrix, with the remaining rows indicating wherever a state coordinate is observed. Our $H$ is independent of time, meaning that we observe the same subset of locations throughout the entire assimilation horizon.

\label{sec:methods}

\subsection{3DVar}\label{sec:3dvar}

We utilize the 3DVar filtering 
algorithm \citep{Lorenc2000} to sequentially estimate the true high-dimensional states $\{x_t^{\text{true}}\}_{t\geq1}$. At each time $t \ge 1$, the state filtering estimate  
$x_{t-1}^s$ is projected forward in time using the
surrogate model dynamics in a forecast step \eqref{eq:forecast}, and this forecast is corrected using the new observation $y_t$ in an analysis step \eqref{eq:analysis}:
\begin{align}\label{eq:forecast}
    \textrm{(forecast)} \ & \ \hat x_{t} = \mathcal{F}_s(x_{t-1}^s), \\
    \textrm{(analysis)} \ & \ x_{t}^s =  \hat x_t + C H^\text{T} (HC H^\text{T} + R)^{-1}(y_t - H\hat x_t). \label{eq:analysis}
\end{align}
For brevity and later reference, we can write \eqref{eq:forecast} and \eqref{eq:analysis} as
\begin{equation}
    x_t^s = (I-KH)\mathcal{F}_s(x_{t-1}^s) + Ky_{t}, \quad t\geq 1 \label{eq:surrogate_3dvar}
\end{equation}
\noindent  where $K= CH^\text{T}(HCH^\text{T}+R)^{-1}$. 
Here, $\mathcal{F}_s$ represents a surrogate forecast model for the dynamics. 
In our experimental results, the observation operator $H$ is described in Section \ref{sec:methodology}\ref{sec:setting}, and the time horizon $T$,  observations $y_t$ for $1 \le t \le T$, and observation error covariance $R$ are described in Section \ref{sec:data}. We next specify the initialization $x_0$, surrogate dynamics map $\mathcal{F}_s,$ and background covariance $C.$  

\paragraph{3DVar initialization $x_0$.} 
We define our 3DVar initialization $x_0$ to be interpolated and standardized $y_0$ data. These low-resolution observations with $\mathcal{N}(0,R)$ additive noise are interpolated to a $720\times 1440$ grid, or a 0.25$^\circ$ resolution, for 20 atmospheric features on January 1, 2023 at 00:00 UTC. The interpolation first uses the nearest neighbor algorithm, then smoothed using a 2D convolution with weight matrix $W^{(k)}\in \mathbb{R}^{k \times k}$, where 
\begin{align}
    W_{i,j}^{(k)} &= \frac{\tilde w_{i,j}}{\sum_i\sum_j \tilde w_{i,j}}, \label{eq:conv_kernel} \\
    \tilde w_{i,j}&=\exp\left\{\frac{-(i-m_i)^2 - (j-m_j)^2}{2\sigma^2}\right\},   \nonumber \\ 
    i &=1,\dots, k, \ j=1,\dots,k,  \nonumber\\ 
    m_i &= \lfloor k/2 \rfloor ,  \ \text{and} \ m_j=\lfloor k/2 \rfloor, \nonumber 
\end{align}
and stride $(1,1)$ for each of the 20 features. For each observation resolution, $\sigma^2=8$.

\paragraph{Surrogate dynamics map $\mathcal{F}_s$.} The surrogate weather model utilized throughout our assimilation experiments takes the form 
 \begin{equation}
     \mathcal{F}_s = S \circ \mathcal{F}_\text{FCN}, \label{eq:F_s}
 \end{equation} 
 where $\mathcal{F}_\text{FCN}$ represents FourCastNet \citep{Pathak2022} and $S$ is a smoothing convolution used to enhance filter stability. 

The Fourier Forecasting Neural Network (FourCastNet) \citep{Pathak2022} provides global weather predictions at 0.25$^\circ$ resolution for short to mid-range time horizons across 20 atmospheric features across various layers of the atmosphere. 
Since FourCastNet combines transformers \citep{Dosovitskiy2021a} and adaptive Fourier Neural Operators \citep{Guibas2021}, evaluating FourCastNet is substantially faster than simulating physics-based weather models, allowing for extremely quick predictions and cheap downstream analysis. 

A known limitation of FourCastNet is its forecasting instability near the poles \citep{Bonev2023}. To enhance filter stability, we utilize a smoothing operator $S$  defined as a 2D convolution with weight matrix $W^{(4)}\in \mathbb{R}^{4\times 4}$ as in equation \eqref{eq:conv_kernel}, and stride $(1,1)$ for each of the 20 atmospheric features. We set $\sigma^2=8$. This smoothing operation attempts to control instabilities in the dynamics model. Appendix B provides an example visualization showcasing filter divergence in its early stage when smoothing is not applied to FourCastNet's forecasts within 3DVar, assimilating 4.5$^\circ$ observations.

\paragraph{Background error covariance $C$.} We specify that $C = qBB^\text{T}$, where $B$ is a matrix representing 2D convolution with weight matrix $W^{(k)}\in \mathbb{R}^{k\times k}$ defined in equation \eqref{eq:conv_kernel}, and stride $(1,1)$ for each of the 20 atmospheric features. In our experiments, we vary the size of the convolutional kernel across observational data resolutions according to $k$. In our experiments, we choose $q = 0.5/ \sum_{i=1}^k\sum_{j=1}^k \{W^{(k)}_{i,j}\}^2$. The constant 0.5 was heuristically chosen to be a similar magnitude to one-step-ahead forecasting errors for $\mathcal{F}_s$ in the standardized space. For each $W^{(k)}$, we set $\sigma^2=8$. With our choices of $C$ and $H$ for each observation resolution, the matrix $(HCH^\text{T} + R)$ in the analysis step in \eqref{eq:analysis} is diagonal, which avoids $d_y^3$ operations for a matrix inversion and instead computes the inverse of $d_y$ scalars, resulting in substantial computational savings. For example when $k=8$, avoiding the matrix inversion reduces the number of computations from on the order of $10^{16}$ operations to on the order of $10^5$ operations.

This $C$ matrix was constructed mainly to maximize computational efficiency and may lead to some physically unrealistic analyses that cause FourCastNet predictions to degrade. Future work can include a more sophisticated construction of this $C$, for example, via the widely adopted National Meteorological Center's method described in \citet{Parrish1992}.

\subsection{Theoretical long-time accuracy of 3DVar}

We are interested in applications where evaluating the ground truth dynamics map $\mathcal{F}$ is unfeasible or computationally expensive, such as a NWP model, but we have a surrogate model $\mathcal{F}_s$ that can be cheaply evaluated, such as FourCastNet as used in \eqref{eq:F_s}. We prove long-time accuracy for a filtering algorithm that uses the
surrogate dynamics $\mathcal{F}_s$ rather than the true dynamics $\mathcal{F}$ in a 3DVar data assimilation task. The result we show relies on (1) standard observability conditions on the true dynamics $\mathcal{F}$ and observation model $H$ and (2)
accuracy of the surrogate model $\mathcal{F}_s$ in the part of the state-space that is not informed by the observations. Here, $\mathcal{F}$ and $\mathcal{F}_s$ represent the flow between observation time points, i.e., 6-hour forecasts. Therefore, it is reasonable to assume that $\mathcal{F}_s$ is a good approximation of $\mathcal{F}$ since surrogate weather models provide accurate short-term predictions.

Formally, the goal of 3DVar is to estimate a signal $\{x_t^\text{true}\}_{t\geq 1}$ given observations $\{y_t\}_{t\geq 1}$ in the setting in \eqref{eq:true_dynamics_obs}.
We want to study the filter accuracy for a surrogate 3DVar filter of the data assimilation scheme defined in \eqref{eq:forecast} and \eqref{eq:analysis}.

\newtheorem{assumption}{Assumption}

\begin{assumption}
 
Suppose the observations we collect are noisy, potentially sparse, unbiased measurements of the ground truth state. More precisely, suppose the data $y_t$ in the surrogate algorithm \eqref{eq:surrogate_3dvar} is found from observing a true signal $x_t^\mathrm{true}$ given by
\begin{align*}
    x_t^\mathrm{true} & = \mathcal{F}(x_{t-1}^\mathrm{true}), \\
    y_t &= Hx_t^{\mathrm{true}} + \gamma \eta_t,
\end{align*}
for $t \geq 1$ and where $\eta_t$ are \textit{i.i.d.}\ and $\mathbb{E}\lVert\eta_t\rVert < A$ for some constant $A>0$. \label{assumption:obs}
\end{assumption}

\newtheorem{theorem}{Theorem}

\begin{theorem}
Suppose Assumption \ref{assumption:obs} holds. Additionally suppose that the Kalman gain matrix $K$ in \eqref{eq:surrogate_3dvar} satisfies that, for some constant $\lambda \in (0,1),$
\begin{equation}
    \lVert(I-KH)D\mathcal{F}(x) \rVert \leq \lambda \quad \forall x \in \mathbb{R}^{d_x},\label{assumption:bound_deriv}
\end{equation}
where $D\mathcal{F}$ denotes the Jacobian matrix of $\mathcal{F}$. Suppose further that 
\begin{equation}
    \lVert(I-KH)(\mathcal{F}_s(x) - \mathcal{F}(x)) \rVert \leq \varepsilon \quad \forall x \in \mathbb{R}^{d_x}.\label{assumption:bound_surrogate}
\end{equation}
Then, there exists a constant $c>0$ independent of $\gamma,$ $\lambda,$ and $\epsilon$ such that the surrogate 3DVar algorithm satisfies 
\begin{equation}
    \lim_{t\to\infty}\sup \mathbb{E}\lVert x_t^s - x_t^{\mathrm{true}} \rVert \leq c\left(\frac{\gamma+\varepsilon}{1-\lambda}\right).\label{eq:upperbound_constant}
\end{equation}

\label{thm:stability}
\end{theorem}

We include a proof of Theorem \ref{thm:stability} in Appendix D.

A key emphasis on our theoretical result is that we only assume that the surrogate model is accurate for short-term horizons, yet we can still obtain long-term analysis stability using it as a dynamics model by leveraging observations $\{y_t\}_{t\geq 1}$. To summarize, our theory rigorously shows that if we have  long-term filter accuracy with the true dynamics model $\mathcal{F}$ and a surrogate model $\mathcal{F}_s$ that provides accurate short-term forecasts, we can achieve long-term filter accuracy with the surrogate dynamics.

\section{Results}\label{sec:results}

Our results evaluate the empirical long-term assimilation stability of 3DVar with our chosen surrogate $\mathcal{F}_s$ in \eqref{eq:F_s} and with varying resolutions of noisy ERA5 data as observations. We evaluate the forecasting performance using our 3DVar analyses as initial conditions and compare against a more naive approach of forecasting using only the interpolated observations as initial conditions. 
These interpolated observations are constructed in the same way as our 3DVar initialization $x_0$ in Section \ref{sec:methodology}\ref{sec:3dvar} for each $\{y_t\}_{t=1}^T$. 
The performance of these two approaches is averaged across 20 standardized atmospheric features and compared to an idealized setting where we compute forecasting metrics using ground truth ERA5 initializations. We include this setting to contextualize how well we could expect to perform in these forecasting tasks in an ideal setting: ground truth information upon initialization. To further explore the task of forecasting, we assess the forecasting performance of an extreme event, Category 5-equivalent super typhoon, Typhoon Mawar in 2023.

We evaluate each of our tasks on 2023 ERA5 reanalysis data using the metrics latitude-weighted root-mean-square-error (RMSE) and latitude-weighted anomaly correlation coefficient (ACC) on ERA5's native 0.25$^\circ$ resolution. We provide a detailed explanation of our error metrics in Appendix E.

\begin{figure*}
  \centering
  \subfloat[a][RMSE across a year-long assimilation window for different observational dataset resolutions, evaluated using standardized 2023 ERA5 data.]{\includegraphics[width=16cm]{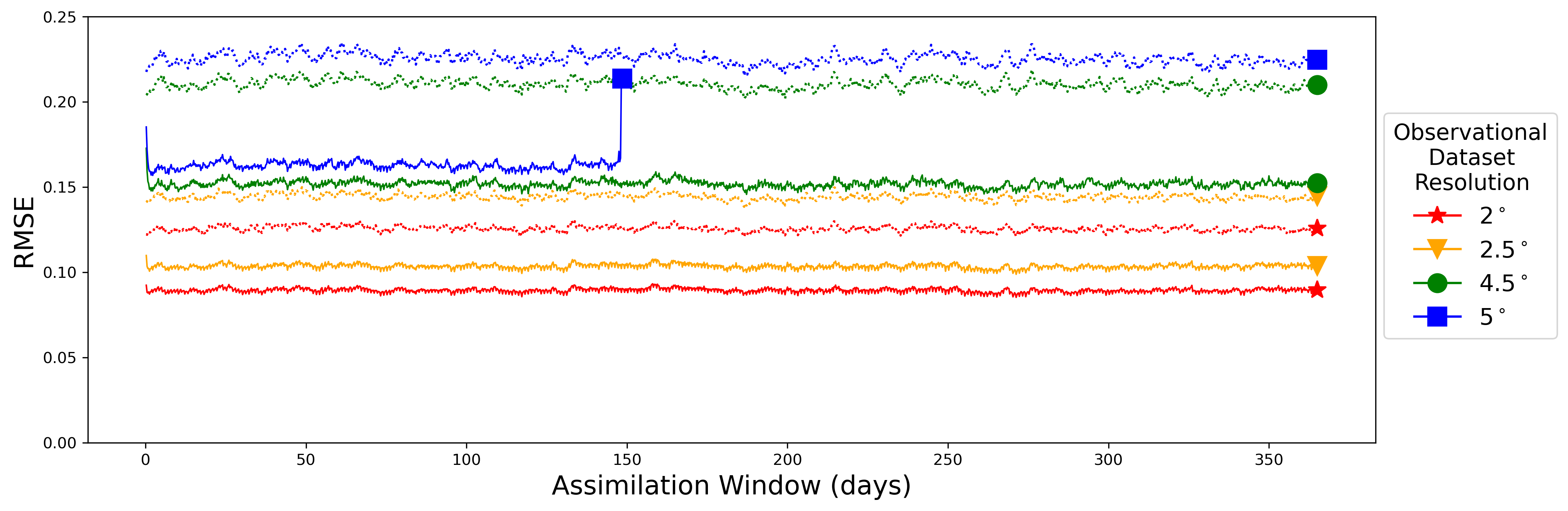}\label{fig:assim_rmse}}  \\
  \subfloat[b][ACC across a year-long assimilation window for different observational dataset resolutions, evaluated using standardized 2023 ERA5 data.]{\includegraphics[width=16cm]{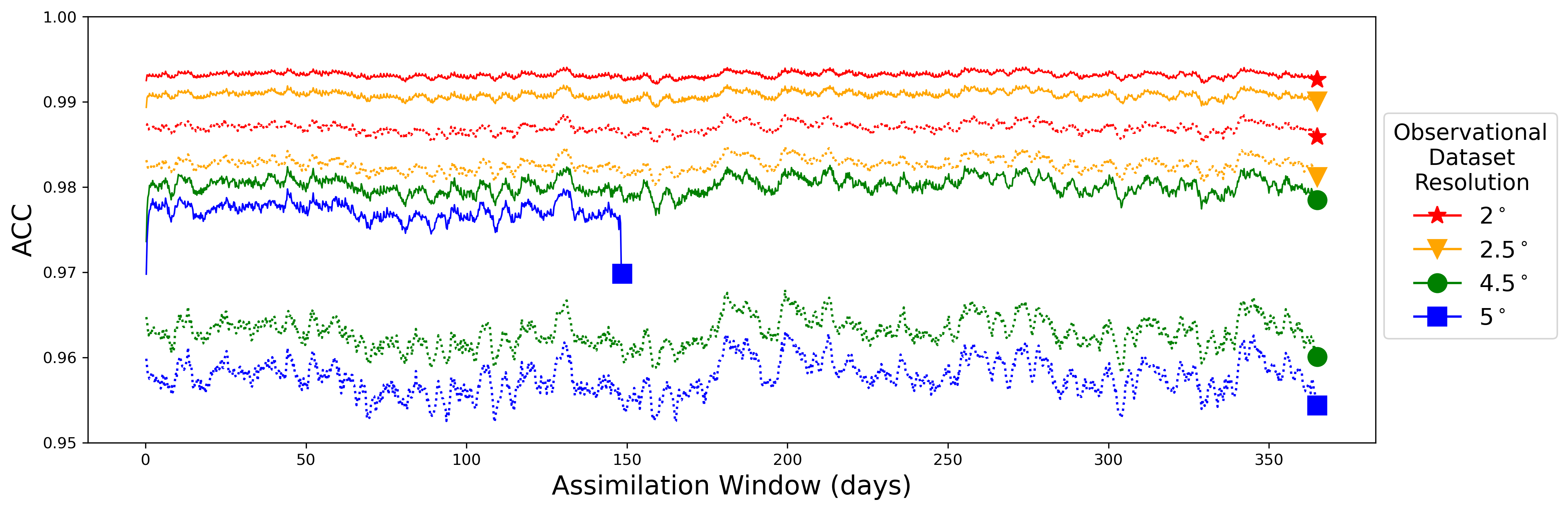}\label{fig:assim_acc}}  \\
  \caption{The dotted lines in both (a) and (b) correspond to metrics for interpolated noisy observations at each time point, and solid lines correspond to metrics for the 3DVar analyses. These metrics are computed using standardized ERA5 data and standardized predictions, and the results are reported as average standardized errors across our 20 atmospheric features. These results show that our 3DVar analyses yield lower RMSE and higher ACC metrics across a year compared to interpolating raw observations. Furthermore, our 3DVar analyses using low-resolution observations achieve stable metrics up to a 5$^\circ$ resolution. At the $5^\circ$ observation resolution, the analysis can be unstable, and we display metrics only up to the time that the instability was detected. 
  } 
\end{figure*}

\subsection{Empirical stability of 3DVar paired with FourCastNet for various observation sizes}\label{sec:vary_obs}

We considered four observation resolutions for our assimilation tasks, ranging considerably in sizes. 
Specifically, our four datasets contain observations of all 20 atmospheric features at every 2$^\circ$, 2.5$^\circ$, 4.5$^\circ$, and 5$^\circ$ in the latitude and longitude directions, with additive $\mathcal{N}(0,R)$ noise. Table \ref{tab:observation_resolutions} provides further details with characteristics about these datasets. We again emphasize that the observation locations remain static throughout our assimilation.

In Figures \ref{fig:assim_rmse} and \ref{fig:assim_acc}, we show the filtering RMSEs and ACCs of our 3DVar implementation for various observation resolutions computed based on ground truth ERA5 data across 2023. As a baseline for comparison, we compute the error for interpolating our low-resolution, noisy observations based on the ground truth ERA5 data. A sufficiently well-calibrated 3DVar implementation would provide better performance compared to this naive baseline, which is the case with our 3DVar analyses. 

\begin{figure*}
  \centering
  \includegraphics[width=16cm]{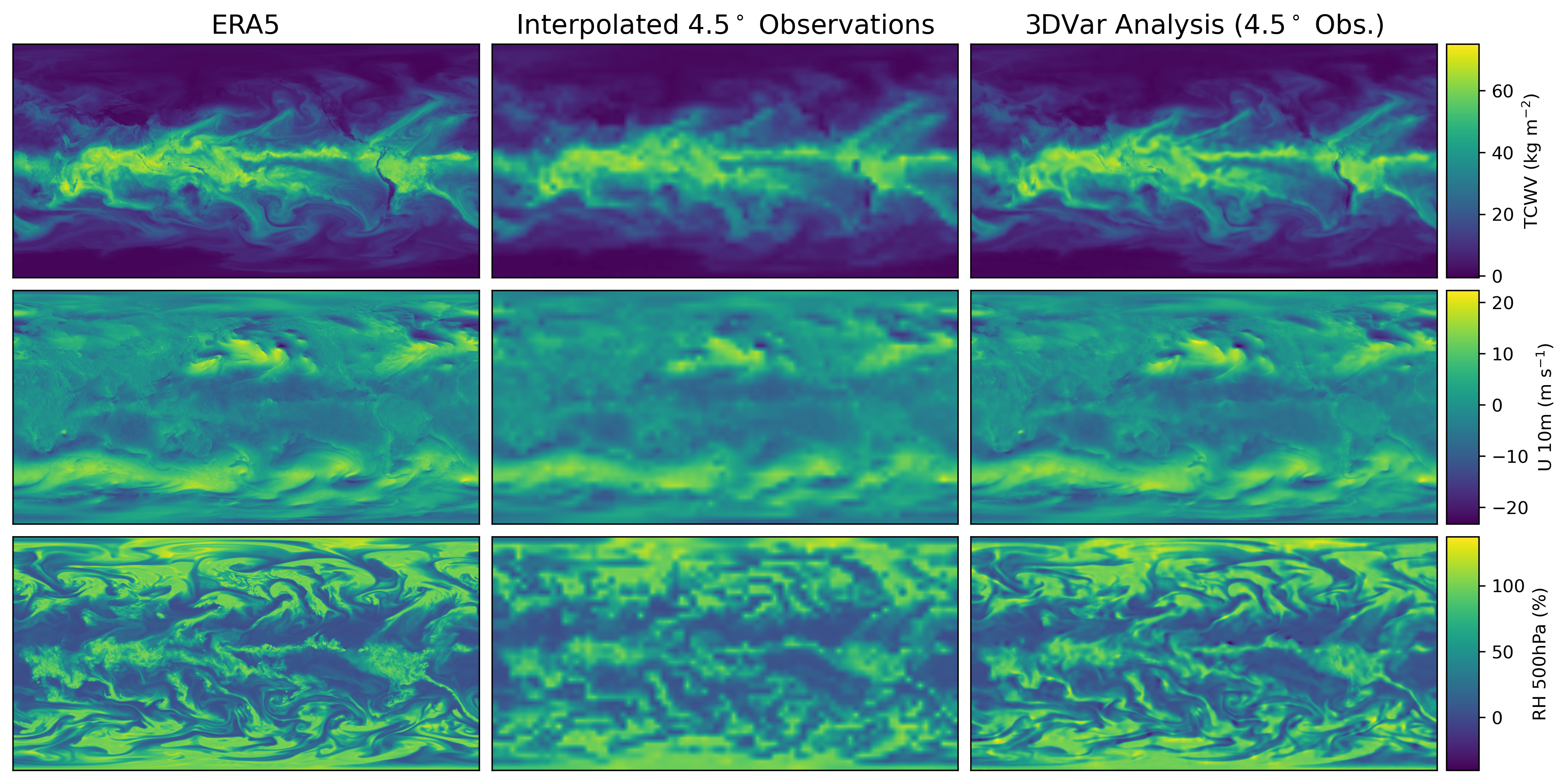}
  \caption{Visualization of the ground truth ERA5 data, interpolated $4.5^\circ$ ERA5 observations with standardized $N(0,0.0001 I_{d_y})$ distributed additive errors, and our 3DVar analysis using this observational data and FourCastNet for the atmospheric features total column water vapor (TCWV), U-component wind speed at 10m above the surface (U 10m), and relative humidity at 500 hPa (RH 500hPa) at the end of our assimilation horizon, December 31, 2023 at 18:00 UTC.} \label{fig:example_assim_result}
\end{figure*}

Comparing our 3DVar analysis RMSEs and ACCs (solid lines) against our naive observation interpolation baseline (dotted lines) for each observation resolution, we notice a consistent and substantial gap in performance in favor of our 3DVar analysis. To qualitatively visualize this gap in performance, we show in Figure \ref{fig:example_assim_result} the ground truth ERA5, interpolated 4.5$^\circ$ observations, and our 3DVar analysis with these 4.5$^\circ$ observations at the end of our assimilation window, corresponding to December 31, 2023 18:00 UTC. The interpolated observations clearly show lack of detail and are overly smooth, which is particularly noticeable in features with sharp gradients throughout the globe, such as relative humidity at 500 hPa. In contrast, our 3DVar analysis using FourCastNet and these $4.5^\circ$ observations show higher quality detail with an appropriate smoothness and detail given the feature. The presence of these details can be attributed to smaller-scale information encoded in the FourCastNet forward pass. To further emphasize this performance gap, we include similar visualizations comparing the ground truth ERA5 data, interpolated $4.5^\circ$ observations, and our 3DVar analyses with these $4.5^\circ$ observations for all 20 atmospheric features at the end of our year-long assimilation in Appendix F. 

We note that the assimilation $5^\circ$ observations in Figures \ref{fig:assim_rmse} and \ref{fig:assim_acc} exhibited filter divergence after assimilating about 150 days worth of observational data, despite the smoothing operation we employed in \eqref{eq:forecast} for filter stability. We visualize in Figure \ref{fig:5deg_diverge} in Appendix B the 3DVar estimate of wind speed compared to ERA5 wind speed soon after the assimilation began to exhibit instability, corresponding to May 29, 2023. The instability originates near the eye of Typhoon Mawar, and FourCastNet predicts increasingly larger wind speeds that are not adequately corrected by the sparse 5$^\circ$ observations in 3DVar. Since a $5^\circ$ is a very sparse dataset, corresponding to observing only 0.25\% of the states, 
and additionally given the simplifying assumptions underlying our construction of $C$ in \eqref{eq:analysis}, 
filter divergence is unsurprising in this extreme case. In the context of our stability theory, for this choice of $K$ that depends on $C$, the upper-bound $\lambda$ in \eqref{assumption:bound_deriv} is large. We speculate that with more sophisticated assumptions on the background covariance $C$ paired with more localized observations, analysis stability for this time horizon may be achievable.

\begin{figure*}
  \centering
  \includegraphics[width=13cm]{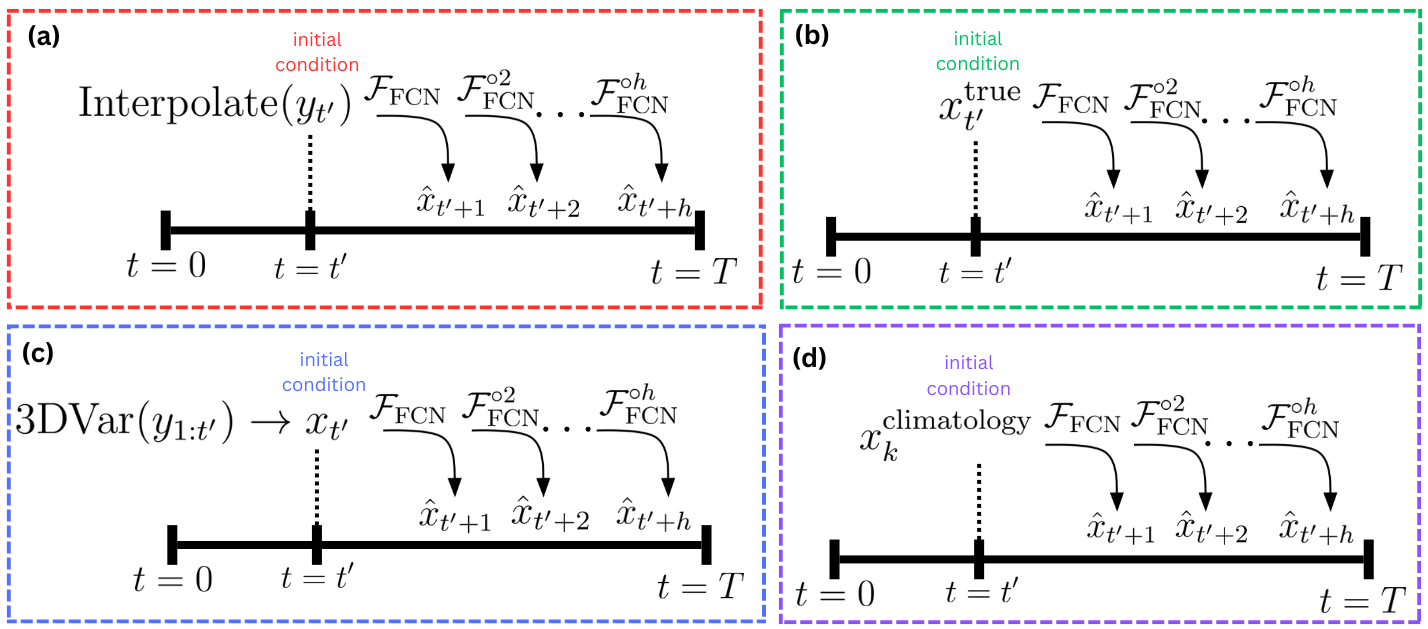}
  \caption{Visualization of various forecasting initializations for the task of $h$-step-ahead forecasting. An initialization at time $t'$ is used to autoregressively  
  compute forecasts up to $h$ time steps ahead using $\mathcal{F}_\text{FCN}$, FourCastNet. The initialization time $t'$ varies between $1\leq t'\leq T-h$ for all tasks (a)-(d). We consider forecasting using (a) interpolated observations, (b) true ERA5 data (unavailable in practice, serving here as an idealized setting), (c) 3DVar analyses, and (d) climatology as initializations. Additionally, $t=0$ corresponds to January 1, 2023 at 00:00 UTC, and $t=T$ corresponds to December 31, 2023 at 18:00 UTC.} \label{fig:forecasting_explanation}
\end{figure*}

\subsection{Forecasting accuracy given various initializations}

We consider the task of $h$-step ahead forecasting given four different types of initializations. More specifically, as shown in Figure \ref{fig:forecasting_explanation} we forecast with (a) interpolated $4.5^\circ$ observations, (b) ground truth ERA5 data (as an ideal setting), (c) 3DVar analyses using these $4.5^\circ$ observations, and (d) climatology as initializations.
Initializations (a) and (d) serve as baselines that a well-calibrated 3DVar analysis would outperform in terms of forecasting error metrics, and (b) serves as a point of comparison in order to tangibly assess the effect of the estimations in (a), (c), and (d). We report metrics in terms of standardized predictions compared to standardized ground truth ERA5 data. We utilize a climatology dataset, which corresponds to the mean value for each spatial location and feature from the years 1979 to 2015 in the ERA5 dataset.

\begin{figure}%
    \centering
    \subfloat[\centering RMSE temporally averaged  across different initializations.]{{\includegraphics[width=7.5cm]{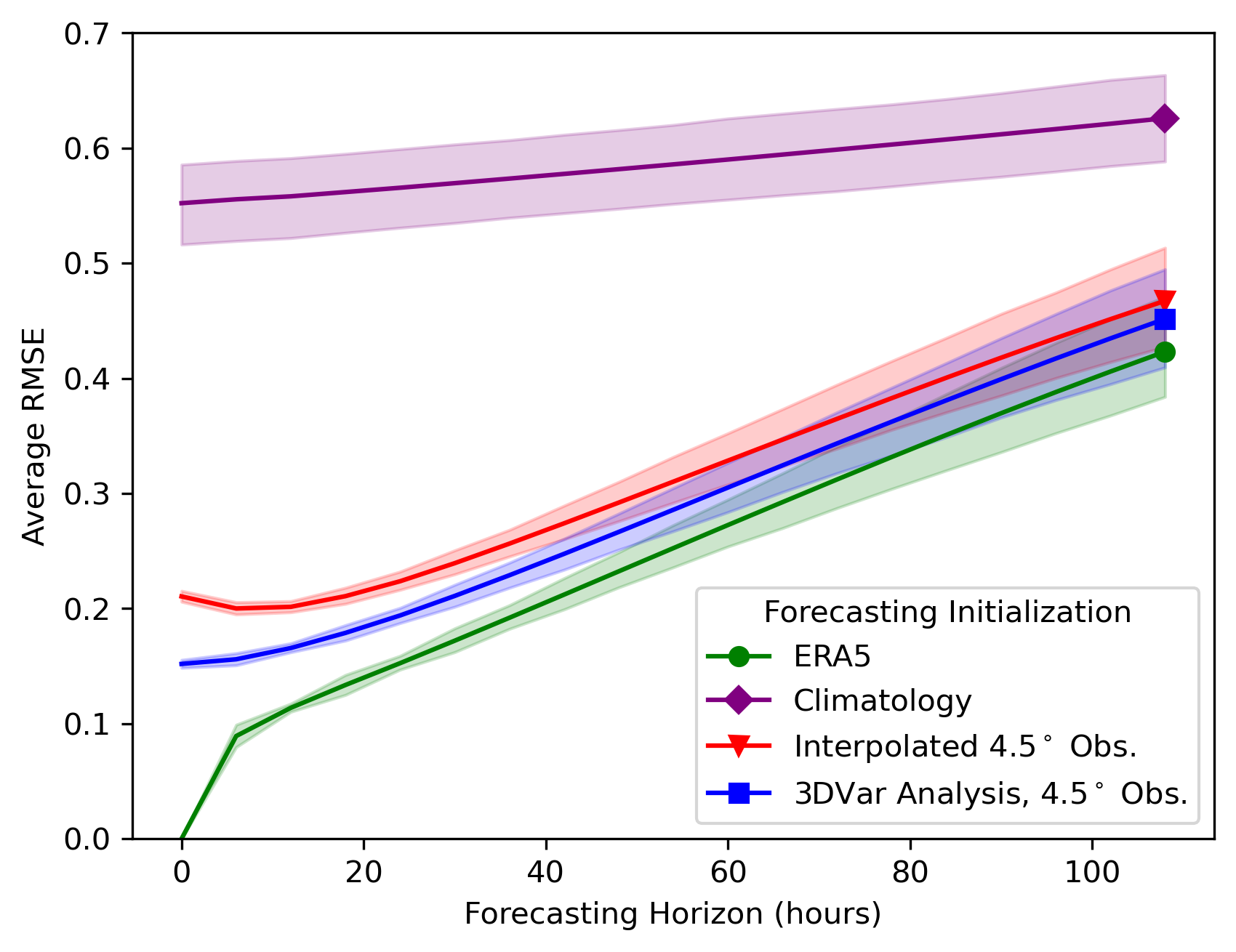} }}%
    \qquad
    \subfloat[\centering ACC temporally averaged across different initializations.]{{\includegraphics[width=7.5cm]{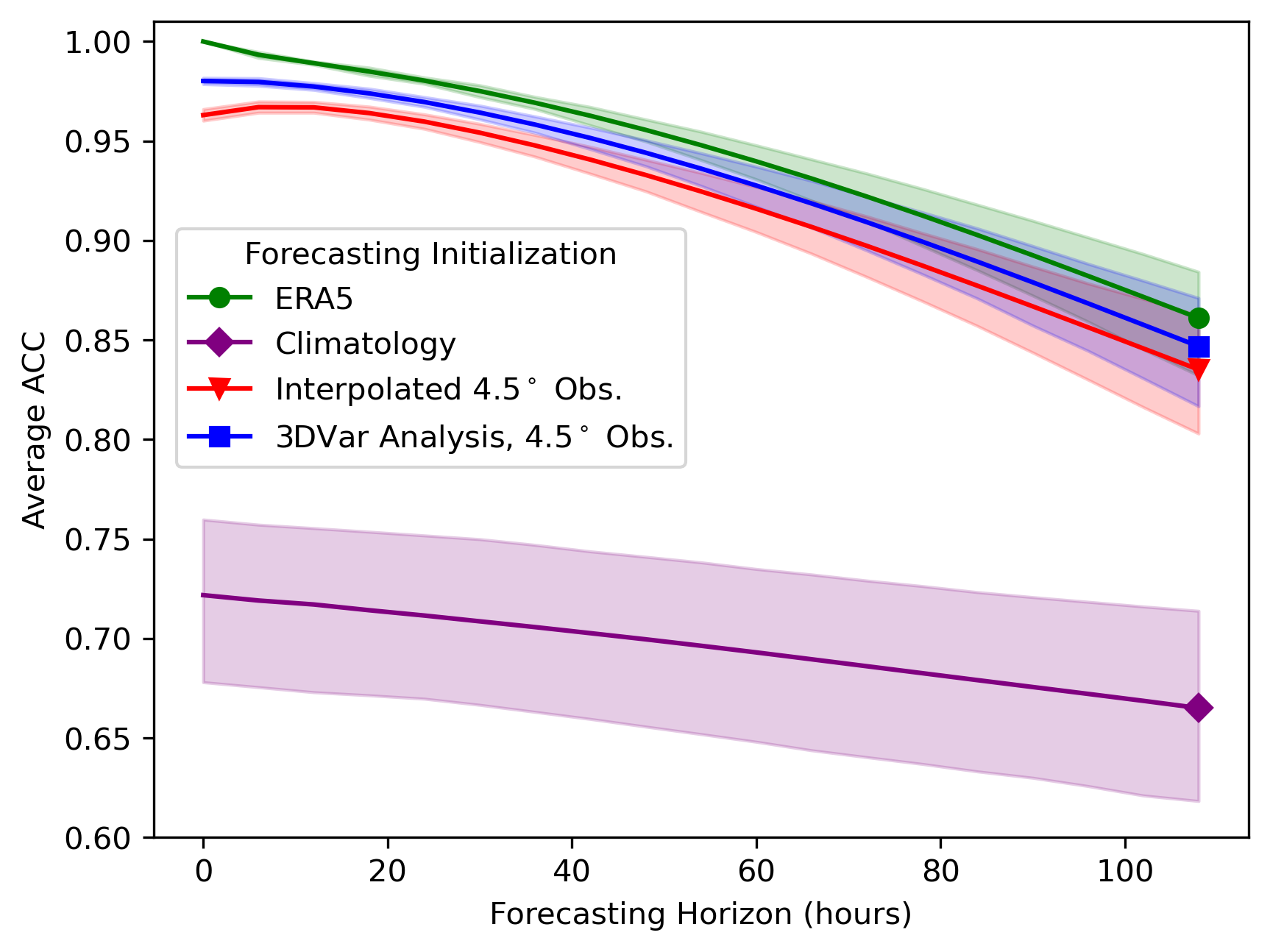} }}%
    \caption{Plots of the 120 hour forecasting performance using (a) interpolated $4.5^\circ$ observations, (b) ground truth ERA5 data, (c) 3DVar analyses with $4.5^\circ$ observation resolution, and (d) climatology as initializations. Each line corresponds to the performance  at each forecasting horizon in 6 hour increments averaged across different time points for the initial conditions. The shaded regions correspond to the 0.05 and 0.95 quantiles of the forecasting metrics at each forecasting horizon. We also plot the $t=0$ errors, which corresponds to the initialization error prior to forecasting.}%
    \label{fig:forecasting_rmse_acc}%
\end{figure}

Figure \ref{fig:forecasting_rmse_acc} shows the RMSEs and ACCs across a 5 day forecasting horizon, averaged across different initial time points within 2023. We note that both the interpolated observations and our 3DVar analyses substantially outperform climatology as an initial condition for our forecasting tasks. This result is to be expected, given that climatology reflects historical averages rather than real-time information that substantially impacts the short-term weather dynamics that we consider. We additionally note that our 3DVar analysis shows a noticeable performance improvement compared to the interpolated observations, particularly in short-term forecasts. The difference in performance is most noticeable within the first roughly 48 hours of the forecast initialization, after which differences in the average forecasting performance become less noticeable.

In producing Figure \ref{fig:forecasting_rmse_acc}, we encountered two out of 1432 time points where our 3DVar analyses with $4.5^\circ$ observations, when used as initial conditions for FourCastNet, resulted in degraded forecasts after a roughly 2 day forecasting horizon. These degraded forecasts are characterized by large, physically unrealistic predicted values originating at a particular location on the globe. The two time points corresponded to an extreme event, specifically Typhoon Khanun, and the forecasting errors autoregressively accumulated in the region of this typhoon. A visualization of the 10m wind speed field across different forecasting horizons that show this divergence is shown in Figure \ref{fig:forecast_diverge} in Appendix B. Such catastrophic forecast errors were not seen in the forecasts using climatology, interpolated $4.5^\circ$ observations, or ground truth ERA5 data.
Because we did not see the same catastrophic forecasting divergence in our 3DVar analyses with $2^\circ$ and $2.5^\circ$ observation resolutions, one hypothesis is that our $4.5^\circ$ 3DVar analyses at these two time points do not have enough data near the typhoon to adequately estimate a physically realistic initial condition given the construction of our 3DVar algorithm, leading to downstream forecasting divergence. However, we also note that some slight perturbation to the locations of assimilated observations leads to stable forecasts for these same two time points. This result suggests that small changes to the assimilated dataset, especially in a highly sparse regime, can lead to large differences in the analyses, and therefore varied forecasts when these analyses are used as initial conditions. Due to the complexity of this system and the black-box nature of machine learning surrogates, a definitive explanation for this behavior is unclear at present.

\subsection{Extreme event: Typhoon Mawar, 2023}\label{subsec:mawar}

Despite the substantial computational advantage data-driven forecasting models provide compared to physics-based models to create forecasting initial conditions, maintaining a satisfactory level of accuracy in forecasts produced from these initial conditions is equally important, especially when considering the substantial impacts that inaccurate forecasts can have on communities during times of extreme events. For example, under-predicting the severity of an extreme event can lead to decision-makers to inadequately inform the public about recommended safety measures. These forecasts need to be accurate enough to properly inform recommendations of disaster mitigation measures, and also computationally cheap enough to be produced in a timely manner.

For these reasons, we narrow our attention in our forecasting evaluation to consider extreme events, and we choose Typhoon Mawar in 2023 as a case study. On May 24, 2023, Typhoon Mawar passed just north of Guam as a category 4-equivalent typhoon, leaving a large portion of the island of 150,000 inhabitants without power \citep{Mawar2023}. Soon after, the typhoon achieved category 5-equivalent status on the Saffir-Simpson Hurricane Wind Scale, with maximum wind speeds recorded on May 26, 2023 \citep{Mawar2023}.

We evaluate FourCastNet's predictions using three different initializations: (1) our 3DVar analysis with $4.5^\circ$ observations, (2) interpolated $4.5^\circ$ observations, and, as an idealized comparison, (3) ERA5 reanalysis data. Comparing our 3DVar-initialized forecasts against interpolated-observation-initialized forecasts allows us to assess the gain in performance as a result of our data assimilation framework, and comparing our 3DVar initialized forecasts against  ERA5 initialized forecasts allows us to assess the performance gap in how well our 3DVar forecasts perform compared to how well we could hope to perform in an ideal scenario where we have access to a high fidelity initial condition. For additional comparison, we include forecasts from (1) the ECMWF's IFS-HRES, which provide high resolution predictions from the IFS numerical weather model, and (2) IBTrACS observational data from the Joint Typhoon Warning Center (JTWC).

\paragraph{Forecasting the eye of Typhoon Mawar from May 23, 2023 00:00 UTC to May 30, 2023 12:00 UTC.}

\begin{figure*}
  \centering
  \includegraphics[width=16cm]{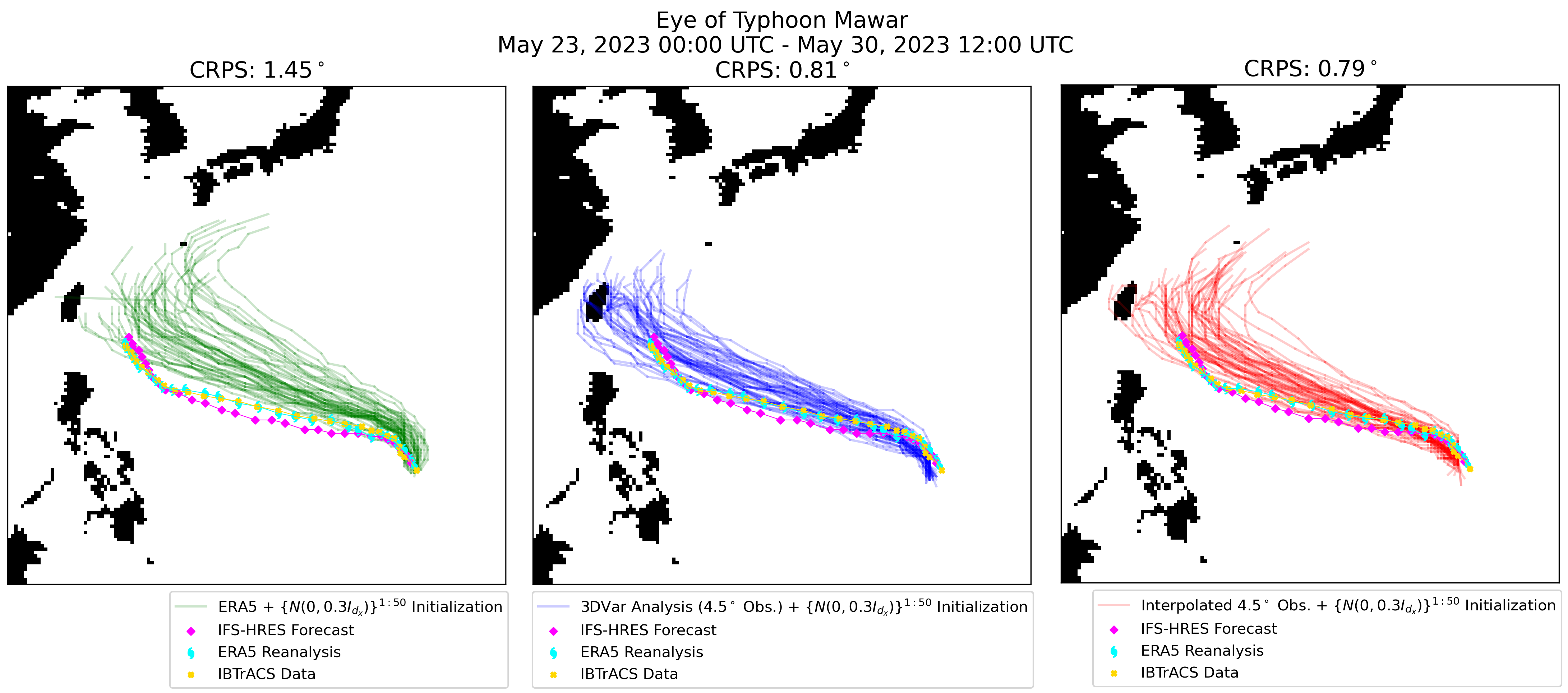}
  \caption{Visualization of FourCastNet's 7 day forecast of the estimated eye of Typhoon Mawar initialized on May 23, 2023 00:00 UTC using three different initial conditions: ground truth ERA5 data as an ideal setting (left), our 3DVar analysis using $4.5^\circ$ noisy observations (middle), and interpolated $4.5^\circ$ noisy observations (right). Each standardized initialization is perturbed by $\mathcal{N}(0,0.3 I_{d_x})$ noise to create a 50 member ensemble. These initial ensemble members were then independently propagated forward in time using FourCastNet without any additional data to correct these forecasts. For comparison, we include the eye of the typhoon based on ERA5, a single IFS-HRES forecast, and IBTrACS observational data in each plot. The skill of the ensemble in predicting the typhoon's trajectory based on IBTrACS data using the CRPS metric is listed at the top of each image showing the forecast ensembles.}
  \label{fig:mawar_estim_eye}
\end{figure*}

Our first typhoon forecasting assessment focuses on the predicted location of the eye of the hurricane, which we characterize by the location of the minimum mean sea level pressure.

For each of our initialization types, we add $\mathcal{N}(0,0.3 I_{d_x})$ noise to each standardized initial estimate and create an ensemble of size 50 with these perturbations. Figure \ref{fig:mawar_estim_eye} visualizes the ensemble of predicted typhoon trajectories for each of our three forecasting initializations. In these plots, we include two trajectories, ``ERA5 Reanalysis'' and ``IBTrACS Data" to evaluate whether these trajectories are included in the ensemble spread for each initialization type. We note that this figure shows forecasts for initializing at only one time point, so the relative performance of initializing with ERA5, our 3DVar analysis, and interpolated observations may vary with different starting time points.

Based on Figure \ref{fig:mawar_estim_eye}, the ensemble spread of the estimated typhoon trajectories generally contain the eye of the typhoon based on the ERA5 reanalysis and IBTrACS data for forecasts using 3DVar analysis (4.5$^\circ$ obs.) and interpolated 4.5$^\circ$ observations as initial conditions. We note that the 3DVar analysis (4.5$^\circ$ obs.) produces a narrower ensemble spread that appears to be closely aligned with these trajectories. However, the FourCastNet predictions initialized with ERA5 reanalysis data appear to be better calibrated with the early-time location of the typhoon's eye; the early-stage forecasts appear to have a slight westward bias for both the 3DVar analysis and interpolated observations initial conditions. The computed continuous ranked probability scores (CRPS) for each ensemble of predictions in Figure \ref{fig:mawar_estim_eye} suggest that the forecasts initialized with the 3DVar analysis (4.5$^\circ$ obs.) and interpolated 4.5$^\circ$ observations are equally performing for predicting the IBTrACS observed typhoon trajectory, and both outperform initializing with ERA5 reanalysis.

All forecasts created using FourCastNet in our visualization share one trait in common: the predictions evolve the typhoon across space at a much faster rate compared to the ground truth. By comparison, the IFS-HRES prediction, despite showing some minor bias throughout the typhoon's trajectory, has a well calibrated speed at which the typhoon moves across the space. As is similarly the case in the ERA5 trajectory, the IFS-HRES forecasts shows a slower initial-time movement, followed by a more rapid north-western movement, then again a slower pace as it dissipates.

\begin{figure}%
    \centering
    \subfloat[\centering Predicted minimum sea level pressure across different initial conditions compared to the actual minimum sea level pressure.]{{\includegraphics[width=7.5cm]{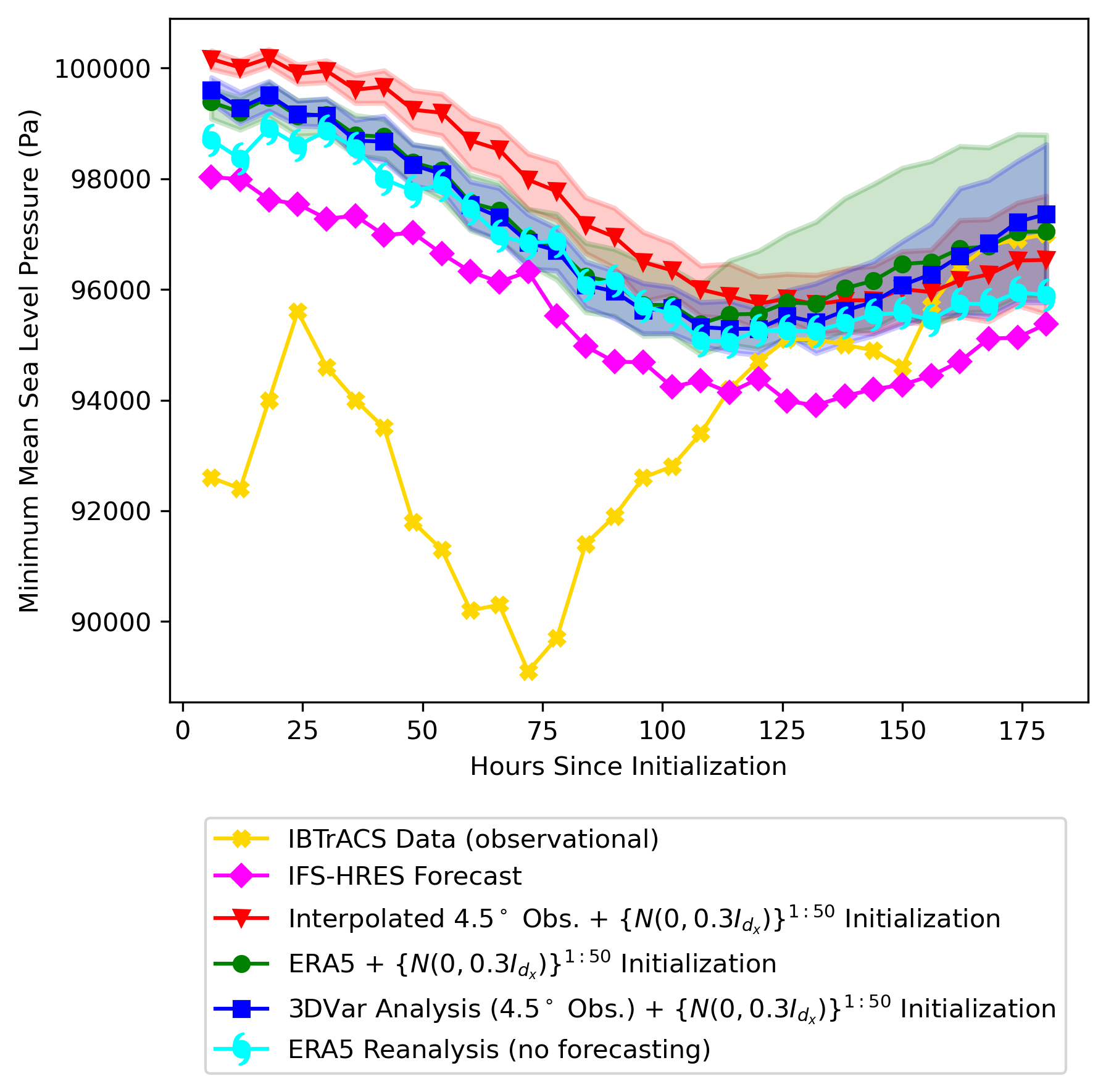} }}%
    \qquad
    \subfloat[\centering Predicted maximum wind speed at 10m above the surface across different initial conditions compared to the actual maximum 10 meter wind speed.]{{\includegraphics[width=7.5cm]{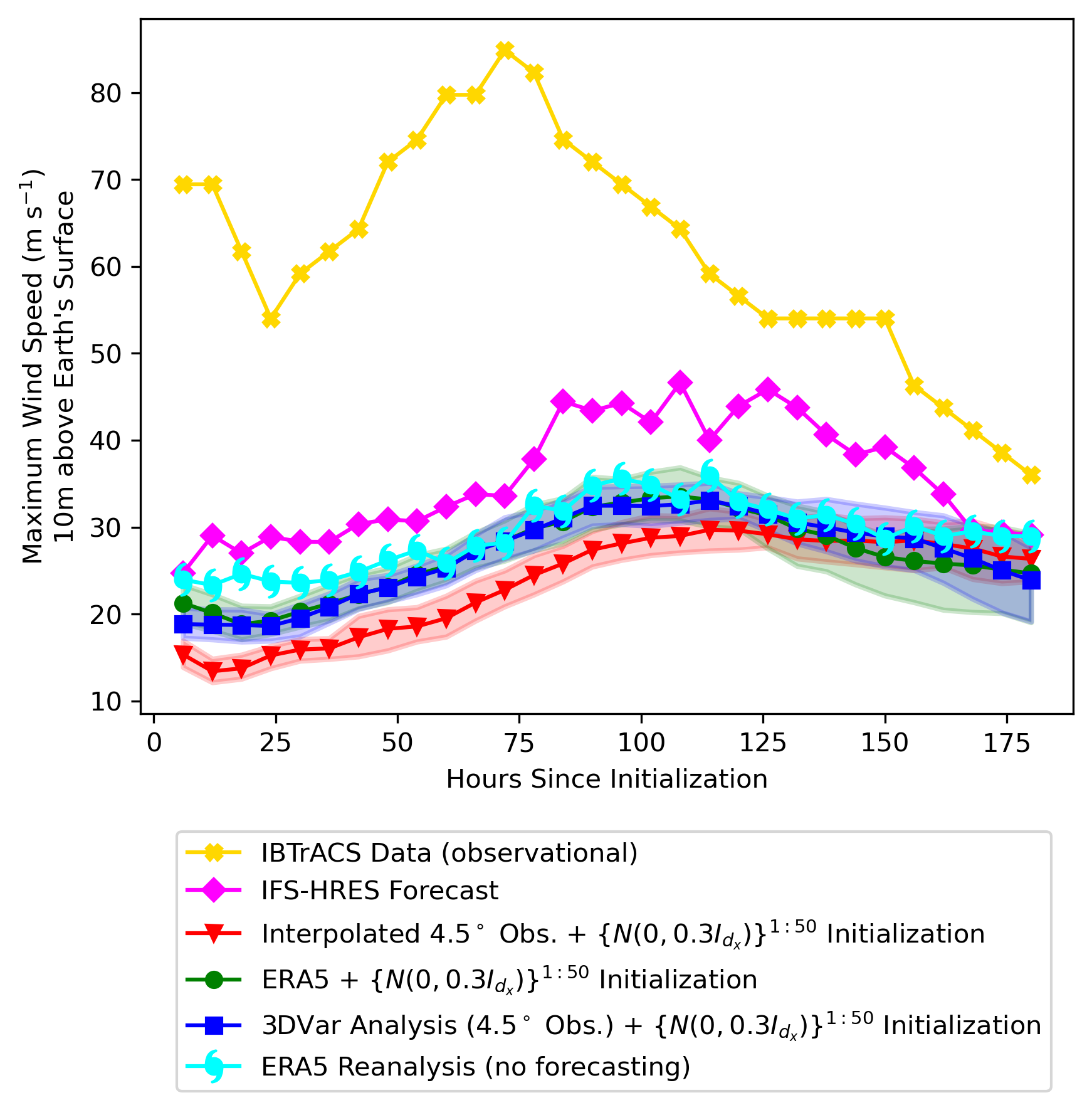} }}%
    \caption{Plots of the forecasted minimum mean sea level pressure and maximum wind speed 10m above the surface for Typhoon Mawar, initialized with ERA5 data, interpolated $4.5^\circ$ observations, and our 3DVar analysis using $4.5^\circ$ observations initialized on May 23, 2023. The ERA5 values, IFS-HRES forecasts, and IBTrACS observations are additionally plotted to compare these forecasts. For each of our three initial conditions, we create an ensemble of size 50 by adding $\mathcal{N}(0, 0.3 I_{d_x})$ distributed noise to the standardized initializations. The forecasts were then converted back to its original scale to produce these plots. The shaded regions in both plots correspond to the 0.05 and the 0.95 quantiles of the ensemble predictions.}%
    \label{fig:mawar_ws_pressure}%
\end{figure}

\paragraph{Forecasting the maximum wind speed 10m above the surface and minimum mean sea level pressure.} Aside from tracking the location of the typhoon, two other important ways to quantitatively characterize the typhoon include assessing the predicted minimum mean sea level pressure and the maximum wind speed at 10m above the surface. These two features determine the categorization of the typhoon, so in order to accurately predict the intensity of the storm, predictions need to be especially accurate regarding these two features. Because wind speed is not a feature native to ERA5 data, we derived the wind speed 10 above the surface using the formula $\sqrt{\text{U10}^2 + \text{V10}^2}$, where U10 and V10 correspond to the U- and V- component wind speed at 10m above the surface, respectively. Figure \ref{fig:mawar_ws_pressure} plots forecasts for these two features across the different initializations we consider, initializing on May 23, 2023 at 00:00 UTC. These plots include the ERA5 reanalysis, IFS-HRES forecast, and IBTrACS observational minimum mean sea level pressure and 10m maximum wind speed across our forecasting horizon as a visual comparison. The IBTrACS maximum wind speeds and minimum pressure values correspond to the most extreme 1-minute sustained values observed.

We note that all three of our initializations, including the IFS-HRES forecast, under-predict the observed intensity of the storm given by the IBTrACS data; the predicted minimum mean sea level pressure is larger than the observed values across our forecasting horizon for all forecasts, and the predicted maximum wind speed at 10m above the surface is lower than the observed value for almost all time points across our forecasting horizon for all three initializations. Both of these results correspond to predicting a less intense typhoon. However, especially apparent in short-time forecasts, our 3DVar analysis produces predicted minimum mean sea level pressure and maximum 10m wind speed closer to the idealized forecasts (i.e., initializing with ground truth ERA5 data) compared to the interpolated $4.5^\circ$ observation forecasts. The smoothed out features in the interpolated $4.5^\circ$ observations likely further contribute to under-predicting this extreme event. As the forecasting horizon increases, the advantage of our 3DVar analysis initialization appears to lessen after an approximate 2 day forecast horizon, as shown by the vast overlap in the ensemble quantiles.

A notable point in Figure \ref{fig:mawar_ws_pressure} is that the ERA5 reanalysis provides significantly less extreme values for the maximum 10m wind speed and minimum mean sea level pressure compared to the IBTrACS observational data, suggesting that the ERA5 may not capture the level of extremity observed for extreme storms.

\begin{figure*}[ht]%
    \centering
    \subfloat[\centering Ground truth ERA5 (left) and 48 hour forecasted (right-most four) wind speed at 10m above the surface on May 25, 2023.]{{\includegraphics[width=14.5cm]{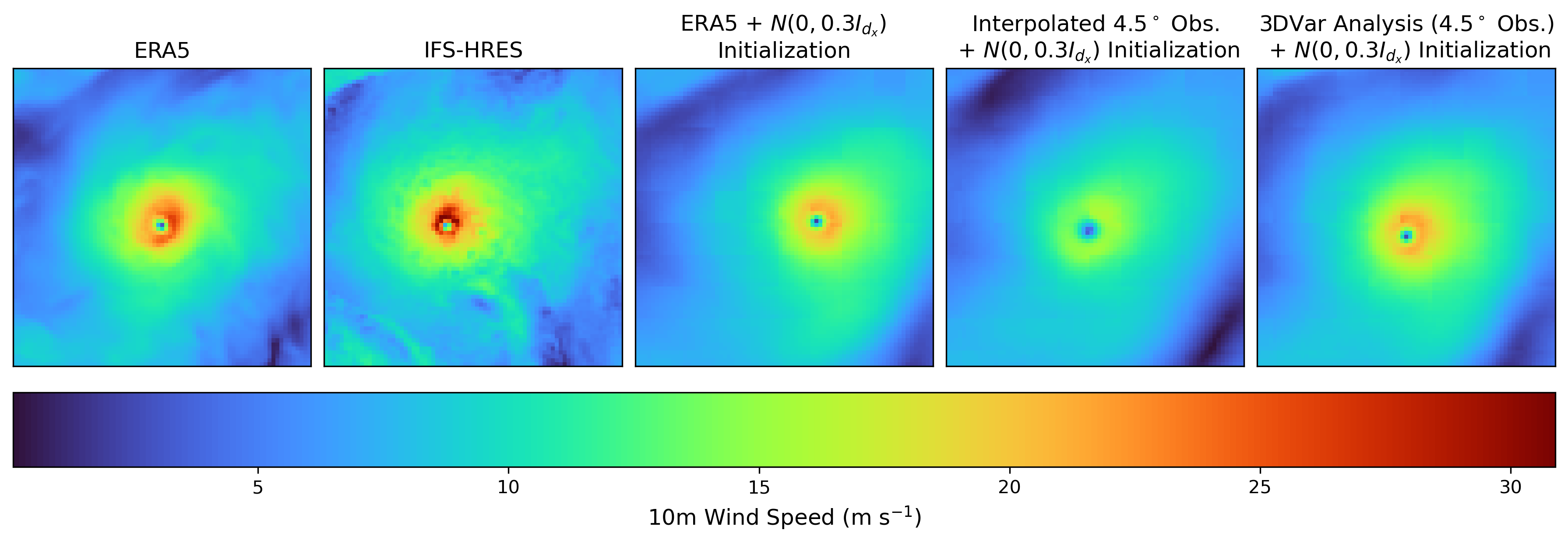} }\label{fig:mawar_ws_one_ens} }%
    \qquad
    \subfloat[\centering Ground truth ERA5 (left) and 48 hour forecasted (right-most four) mean sea level pressure on May 25, 2023.]{{\includegraphics[width=14.5cm]{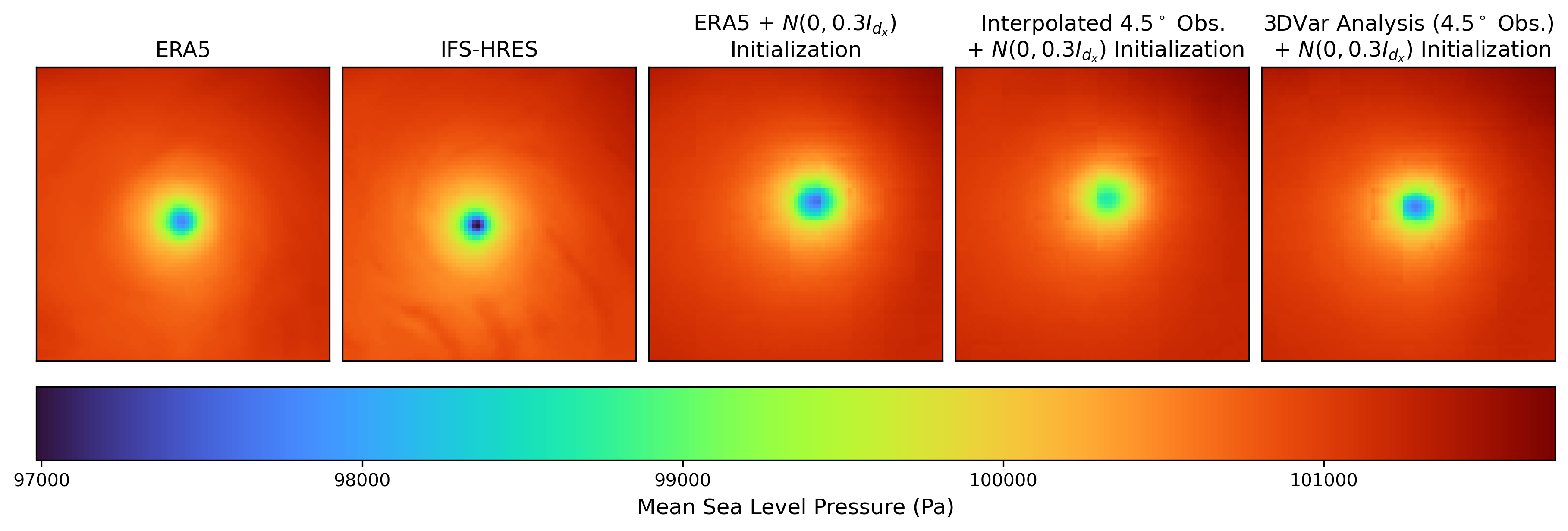} }\label{fig:mawar_pressure_one_ens}}%
    \caption{Visualization of the ground truth ERA5 wind speed at 10m above the surface and mean sea level pressure compared to 48 hour forecasts from IFS-HRES and FourCastNet's forecasts given three initial conditions: noisy ERA5 data, noisy interpolated $4.5^\circ$ observations, and a noisy 3DVar analysis ($4.5^\circ$ observations).}%

\end{figure*}

\paragraph{48 hour forecast of Typhoon Mawar, initialized on May 23, 2023 00:00 UTC.} To qualitatively visualize the difference in predictions for our three initial conditions in terms of the wind speed and the sea level pressure, we provide plots of one ensemble member's 48 hour forecasts in Figures \ref{fig:mawar_ws_one_ens} and \ref{fig:mawar_pressure_one_ens}, corresponding to May 25, 2023 at 00:00 UTC, using our three initializations. We include the ERA5 reanalysis and the IFS-HRES forecast for May 25, 2023 in these figures as a visual comparison. Consistent with the conclusion drawn from Figure \ref{fig:mawar_ws_pressure}, the 48 hour forecast for interpolated $4.5^\circ$ observations shows a less extreme forecast in terms of wind speed and mean sea level pressure compared to the ERA5 reanalysis. The forecasts from initializing with the ERA5 reanalysis and a 3DVar analysis ($4.5^\circ$ observations) show a qualitatively better prediction of the eye of the hurricane, more closely matching its intensity. For a more thorough assessment across different ensemble members, we include a comparison of three other ensemble forecasts in Appendix G. Lastly, we note that Figures \ref{fig:mawar_ws_one_ens} and \ref{fig:mawar_pressure_one_ens} visually reiterate the insights that the FourCastNet predictions under-predicted the intensity of the typhoon, while the IFS-HRES forecast is closer to the extremity observed in the IBTrACS observations, though the forecast intensity still falls short of the observed intensity.

\section{Conclusions}\label{sec:conclusions}

We empirically show and theoretically justify that filter stability, particularly using 3DVar, is achievable in settings with surrogate weather dynamics models, which allow for substantial computational speedup compared to filtering with NWP models. We additionally show that forecasting with 3DVar-based initializations produced from a data-driven weather surrogate can provide more accurate short-term predictions than with more naive approaches to initialization, such as using interpolated observations. We note that results using 3DVar with data-driven weather surrogates can potentially be improved with a more physically-informed choice of background covariance $C$ by providing improved stability.

The success of our filtering experiments in our global setting offers promise to future directions. Specifically, a more challenging yet important future direction is assimilating real observations, which involves data collected irregularly over the globe, and measurements that may not directly correspond to an atmospheric feature of interest. In these settings, a nonlinear observation operator is needed to transform the observations from quantities that indirectly measure the feature of interest into that feature. Future research in this direction can assess issues that arise due to non-linearities in the observation process, as well as the impact of associated measurement noise on the long-term filtering stability.

We emphasize that, despite the fact that our observational dataset (1) lies on a regular grid, (2) directly measures the quantities of interest, and (3) has low measurement error, our results provide positive implications for the task of assimilating coarse NWP forecasts. Assimilating these coarse forecasts can be done to reduce the cost of expensive NWP solvers while still providing high-resolution, accurate estimates of atmospheric states. These coarse NWP forecasts can provide important information to assimilate into surrogate models' forecasts, particularly in maintaining longer horizon forecast accuracy, at a cost substantially cheaper than high-resolution NWP solves. Since our results show physically realistic-looking filtering results that have stable errors over a long time horizon, data assimilation with weather surrogates shows substantial promise for applications in this direction.

\section{Acknowledgments}


MA is grateful to be supported by the National Science Foundation Graduate Research Fellowship under Grant No. DGE-1746045. DSA is grateful for the support of the NSF CAREER award DMS-2237628, DOE DE-SC0022232, and the BBVA Foundation. RW is grateful for the support of NSF DMS-1930049, NSF OAC-1934637, DOE DE-SC0022232, and NSF DMS-2023109. This research used resources of the National Energy Research Scientific Computing Center, a DOE Office of Science User Facility supported by the Office of Science of the U.S. Department of Energy under Contract No. DE-AC02-05CH11231 using NERSC award ASCR-ERCAP0022809.

The authors also thank Mihai Anitescu and Philip Dinenis for helpful discussions regarding our experimental results,  and Jaideep Pathak, Morteza Mardani, and Karthik Kashinath for useful insights in the motivation for this work.

\section{Data Availability}

In this work, we utilized three datasets: (1) the ERA5 reanalysis data, which can be downloaded at the Copernicus Climate Data Store, \citet{ERA5Pressure} for ERA5 data on pressure levels and \cite{ERA5Land} for ERA5 data on land, (2) ECMWF's IFS-HRES forecasts initialized at either 00:00 UTC or 12:00 UTC daily, which can be downloaded at the TIGGE Data Retrieval portal \citep{TIGGE}, and (3) NOAA's International Best Track Archive for Climate Stewardship (IBTrACS) data \citep{IBTrACS, Gahtan2024}.

\bibliographystyle{abbrvnat}
\bibliography{references}

\clearpage
\appendix

\section{Visualization of ground truth ERA5 and 4.5 degrees ERA5 observations for relative humidity at 500 hPa.}\label{app:obs_vs_latent}

\begin{figure*}[ht]%
    \centering{\includegraphics[width=14.5cm]{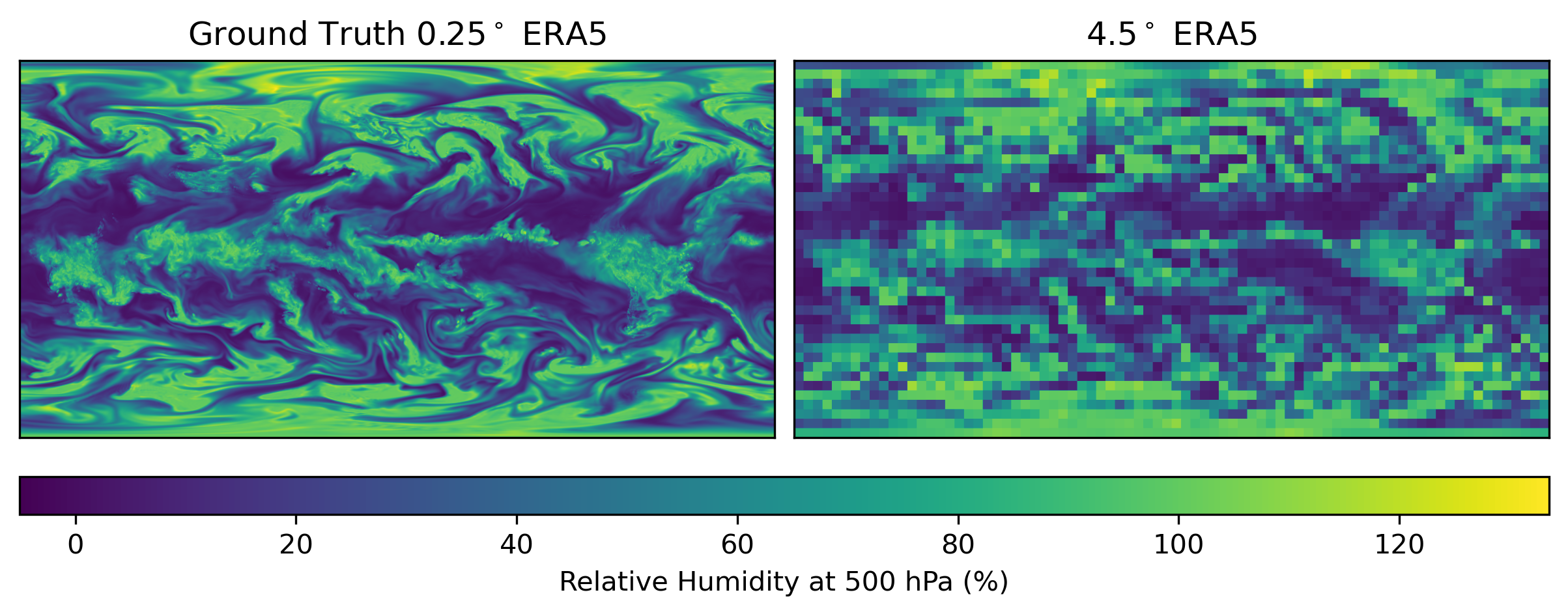}} %
    
    \caption{Visual comparison of $0.25^\circ$ ERA5 data and our $4.5^\circ$ observations created by systematically thinning out the $0.25^\circ$ ERA5 data. This plot corresponds to December 31, 2023 at 18:00 UTC.}\label{fig:latent_vs_obs} 
\end{figure*}

Figure \ref{fig:latent_vs_obs} provides a visual comparison of one time point of relative humidity at 500 hPa for $0.25^\circ$ ERA5 compared to ERA5 data thinned out to a resolution of $4.5^\circ$. In plotting the 4.5$^\circ$ observations, we mapped this data up to a 720$\times$1440 grid via nearest neighbors, resulting in a pixelated-looking image.

\section{Filter and Forecasting Divergence} \label{app:diverge}
    In this appendix, we visualize instances of divergence that we encountered in our experiments.

\paragraph{Filter divergence without forecast smoothing.}

\begin{figure*}
  \centering
  \includegraphics[width=16cm]{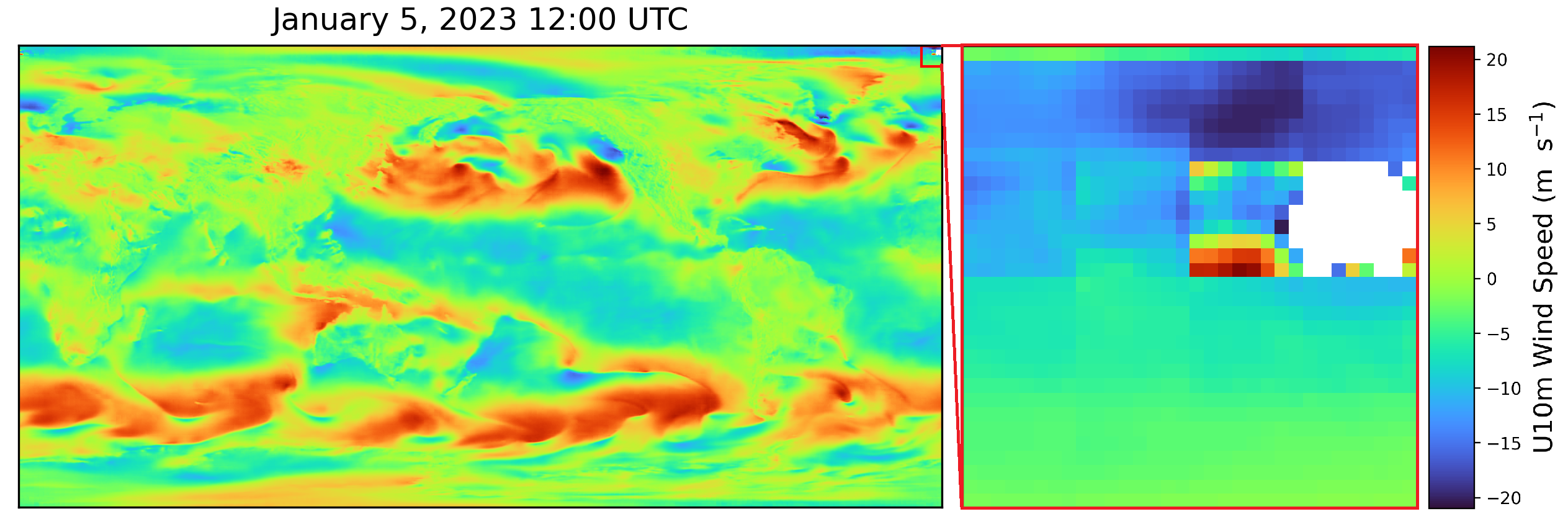}
  \caption{Example visualization of the 3DVar analysis without forecast smoothing (i.e., $\mathcal{F}_s := \mathcal{F}_{\text{FCN}}$) for U-component wind speed 10m above the surface. This 3DVar analysis corresponds to January 5, 2023 12:00 UTC after assimilating 4.5$^\circ$ observations 6-hourly starting January 1, 2023 00:00 UTC. In this visualization, the extreme values that are characteristic of filter divergence are filled in with white pixels. A pixel is characterized as diverging and filled in with white if it was 10\% larger than either the minimum or maximum ERA5 value for that same time point.} \label{fig:no_smoothing_divergence}
\end{figure*}

Figure \ref{fig:no_smoothing_divergence} shows a 10m U-component wind speed estimated state from 3DVar for assimilating 4.5$^\circ$ observations without applying a smoothing convolution on FourCastNet's forecasts, meaning $\mathcal{F}_s := \mathcal{F}_\text{FCN}$ in equation \eqref{eq:F_s}. The analysis state exhibits early-stage filter divergence for this assimilation task near the top right corner of the image. The initial sign of filter divergence appeared on January 5, 2023 at 12:00 UTC, so the filter was only able to assimilate about 4 days of 6-hourly observational data before showing signs of degradation. There is no obvious physical meaning behind the origin of the filter divergence, so it may be reasonable to assume that the 3DVar analysis inputs to FourCastNet are different enough from the ERA5 dataset that instabilities arise, particularly near the northern pole. FourCastNet version 1 is known to degrade near the poles when applied autoregressively \citep{Bonev2023}, so an interesting future direction would be to assess FourCastNet v2's stability in data assimilation tasks.

\begin{figure*}
  \centering
  \includegraphics[width=16cm]{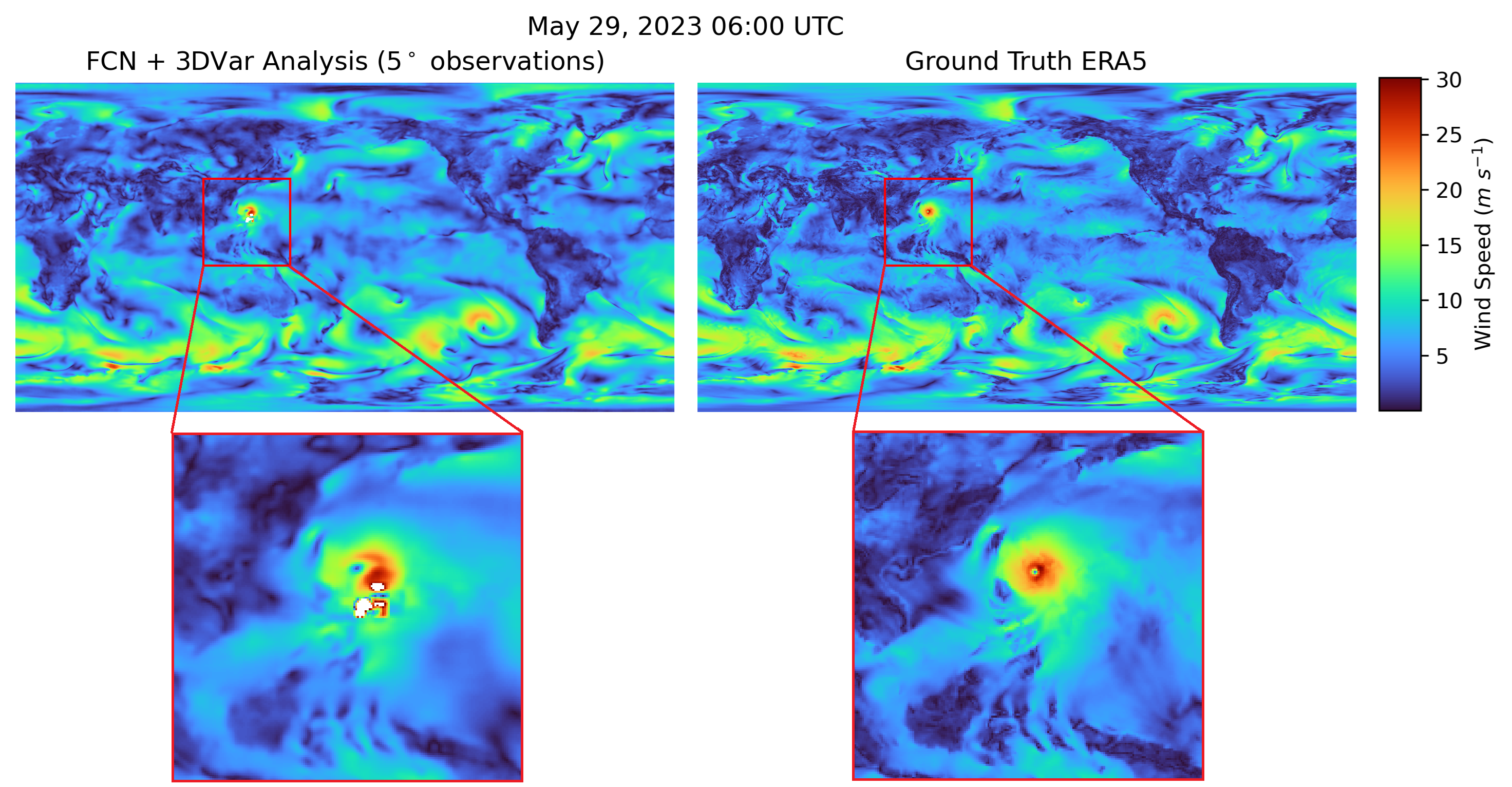}
  \caption{Visualization of the observed beginning of filter divergence for the task of assimilating 5$^\circ$ observations into FourCastNet's predictions compared to the ground truth ERA5. This particular snapshot corresponds to global 10m wind speed on May 29, 2023 at 06:00 UTC. The zoomed in region in the 3DVar analysis and the ground truth ERA5 data localize Typhoon Mawar, which is the origin of the 3DVar filter divergence. The colorbar maps the 10m wind speed from the values 0 to the maximum seen in ERA5, which corresponds to roughly 30 m s$^{-1}$. Extreme values in the 3DVar analysis above 30 m s$^{-1}$ were masked in plotting and mapped to a solid white color. In this particular snapshot, the 3DVar analysis had an estimated maximum wind speed of roughly 50 m s$^{-1}$, which is almost double the value seen in ERA5.} \label{fig:5deg_diverge}
\end{figure*}

\paragraph{Filter divergence for assimilating 5$^\circ$ observations.} Here, we further investigate the filter divergence for assimilating 5$^\circ$ observations seen in Figures \ref{fig:assim_rmse} and \ref{fig:assim_acc}. Figure \ref{fig:5deg_diverge} visualizes the 3DVar estimate of the state for 10m wind speed on May 29, 2023, corresponding to roughly two days before filter divergence is clearly seen in the MSE and ACC metrics. As shown in Figure \ref{fig:5deg_diverge}, the filter divergence originates at the center of an extreme event, specifically Typhoon Mawar. The filter divergence originating from this extreme event suggests that, given our assumptions on $C$ within 3DVar, 5$^\circ$ observations do not provide enough localized information around the typhoon to stabilize FourCastNet's predictions, and these unstable predictions are not adequately corrected within 3DVar based on our static constructing of $C$. As an additional note, we see that the 3DVar analysis lacks the expected small-scale structure of the eye of the typhoon: the largest wind speed values occur in multiple modes rather than as a ring around a low wind speed eye, which further evidences the unphysical nature underlying the divergence.

\begin{figure*}
  \centering
  \includegraphics[width=16cm]{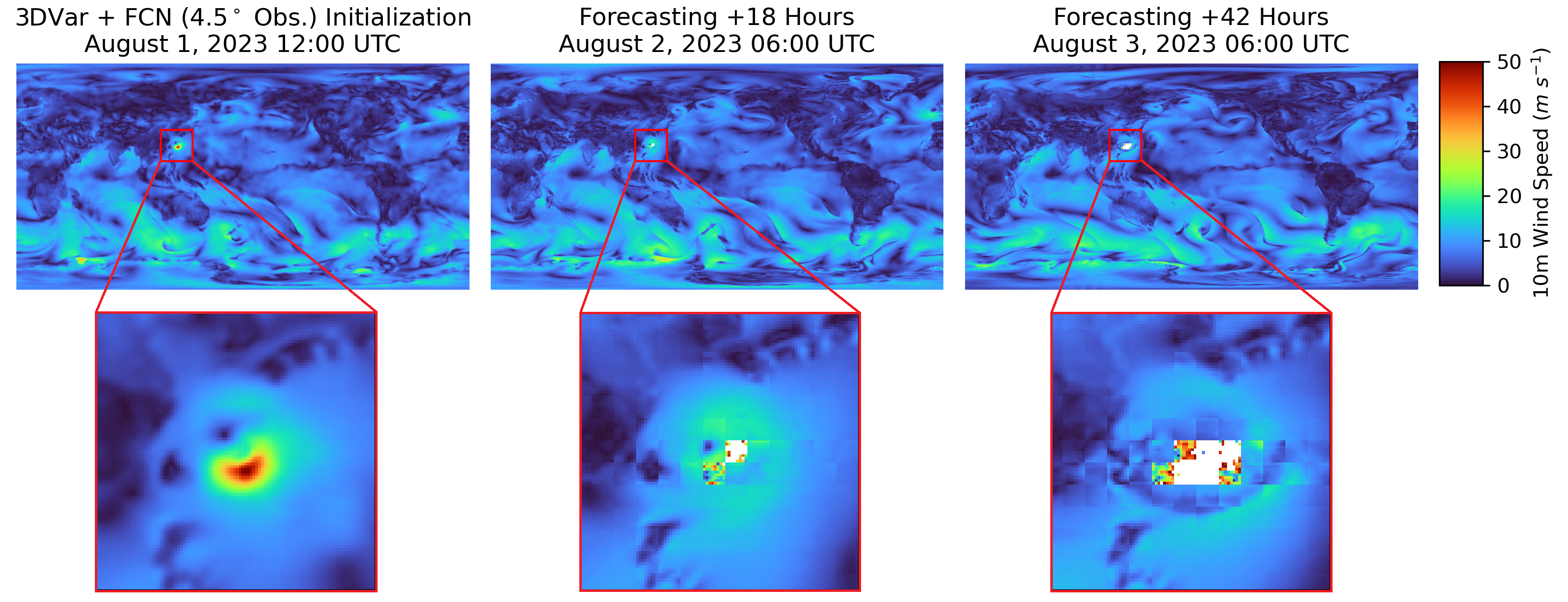}
  \caption{Visualization of an example 3DVar analysis (4.5$^\circ$ observations) of 10m wind speed that, when used as an initialization for FourCastNet, rapidly leads to a divergent forecast. The left-most image shows the 3DVar analysis (4.5$^\circ$ observations) initialization prior to forecasting with FourCastNet. The middle image shows the predicted global 10m wind speed at a 18 hour forecast horizon, and the right-most image shows the predicted global 10m wind speed at a 42 hour forecast horizon. Extreme predicted values are masked by white pixels. A pixel was determined as extreme if it exceeded the maximum wind speed in the initialization, which is roughly 50 m s$^{-1}$.} \label{fig:forecast_diverge}
\end{figure*}

\paragraph{Forecast divergence for initializing forecasts with 3DVar analyses (4.5$^\circ$ observations).} We encountered two particular instances of forecast divergence when initializing with our 3DVar analyses that assimilate 4.5$^\circ$ observations. In particular, initializing on August 1, 2023 at 12:00 UTC and 18:00 UTC lead to forecast divergence within a 24 hour forecast horizon. Figure \ref{fig:forecast_diverge} shows that when initializing forecasts on August 1, 2023 at 12:00 UTC, the forecast divergence originates at the center of a typhoon in the Pacific Ocean, specifically Typhoon Khanun. The visualization for forecasting with a 3DVar analysis (4.5$^\circ$ obs.) on August 1, 2023 at 18:00 UTC shows similar behavior to Figure \ref{fig:forecast_diverge}. 

The 3DVar analyses (4.5$^\circ$ obs.) for these two time points notably have estimated wind speeds much larger than is seen in the ERA5 dataset: about 50 m s$^{-1}$ for the 3DVar analyses and about 30 m s$^{-1}$ for ERA5. The large overestimate in the initial condition causes instabilities in the forecast that quickly propagate, as is shown in Figure \ref{fig:forecast_diverge}.

\section{Background Error Covariance} \label{app:background_cov}

\begin{figure*}
  \centering
  \includegraphics[width=12cm]{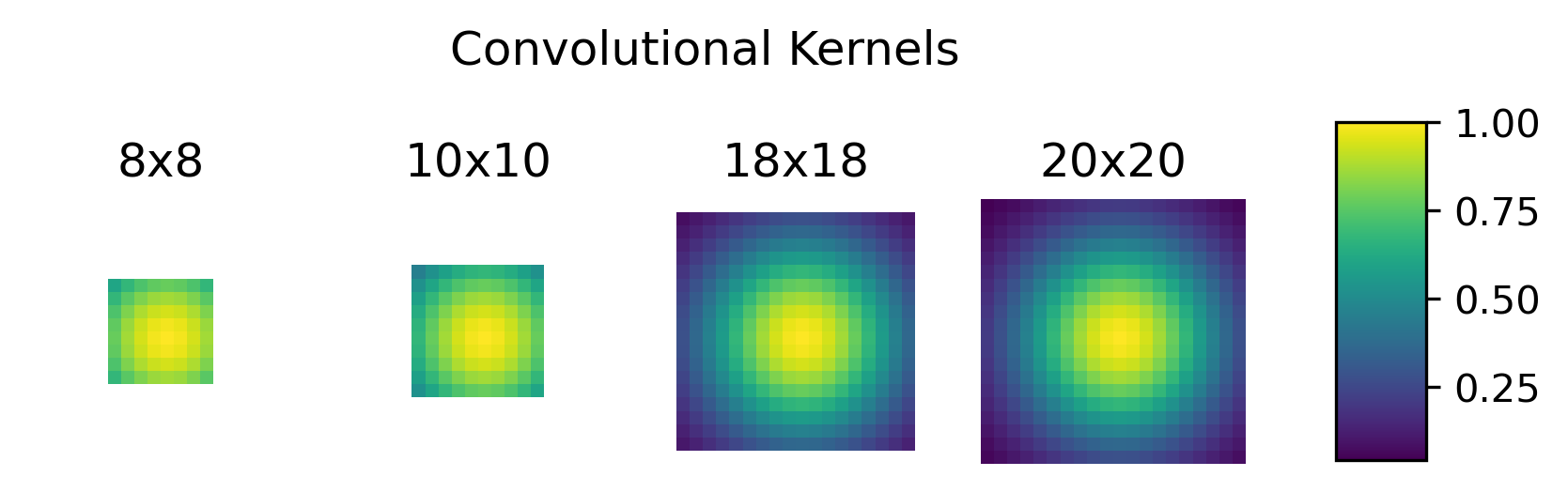}
  \caption{Visualization of the different kernels $W^{(k)}$ from equation \eqref{eq:conv_kernel} for $B$ utilized in our data assimilation tasks for $k=8,10,18,$ and $20$. Each kernel has the shared parameter $\sigma^2=8$ in defining the Gaussian decay.} \label{fig:conv_kernels}
\end{figure*}

We construct our background error covariance with simplified assumptions that lead to computational efficiencies in 3DVar. In particular, we assume that the covariance between two distinct locations is proportional to a truncated Gaussian decay as described in equation \eqref{eq:conv_kernel} and visualized in Figure \ref{fig:conv_kernels}. This construction assumes that the covariance between two distinct locations decreases as the distance between the two locations increases, and when the distance between the two points reaches a certain threshold ($k/2$ in our experiments), the covariance immediately drops to 0. In all four kernels, the scaling parameter $\sigma^2$, which controls how quickly the association decays as the distance between two locations increases, is held constant at 8 for each setting. The parameter $\sigma^2=8$ was chosen so that the covariance decays gradually within our chosen convolutional kernel size, as shown in Figure \ref{fig:conv_kernels}.  The main difference among the four kernels is the distance at which the covariance becomes 0, which is dependent on $k$.

\begin{figure*}
  \centering
  \includegraphics[width=12cm]{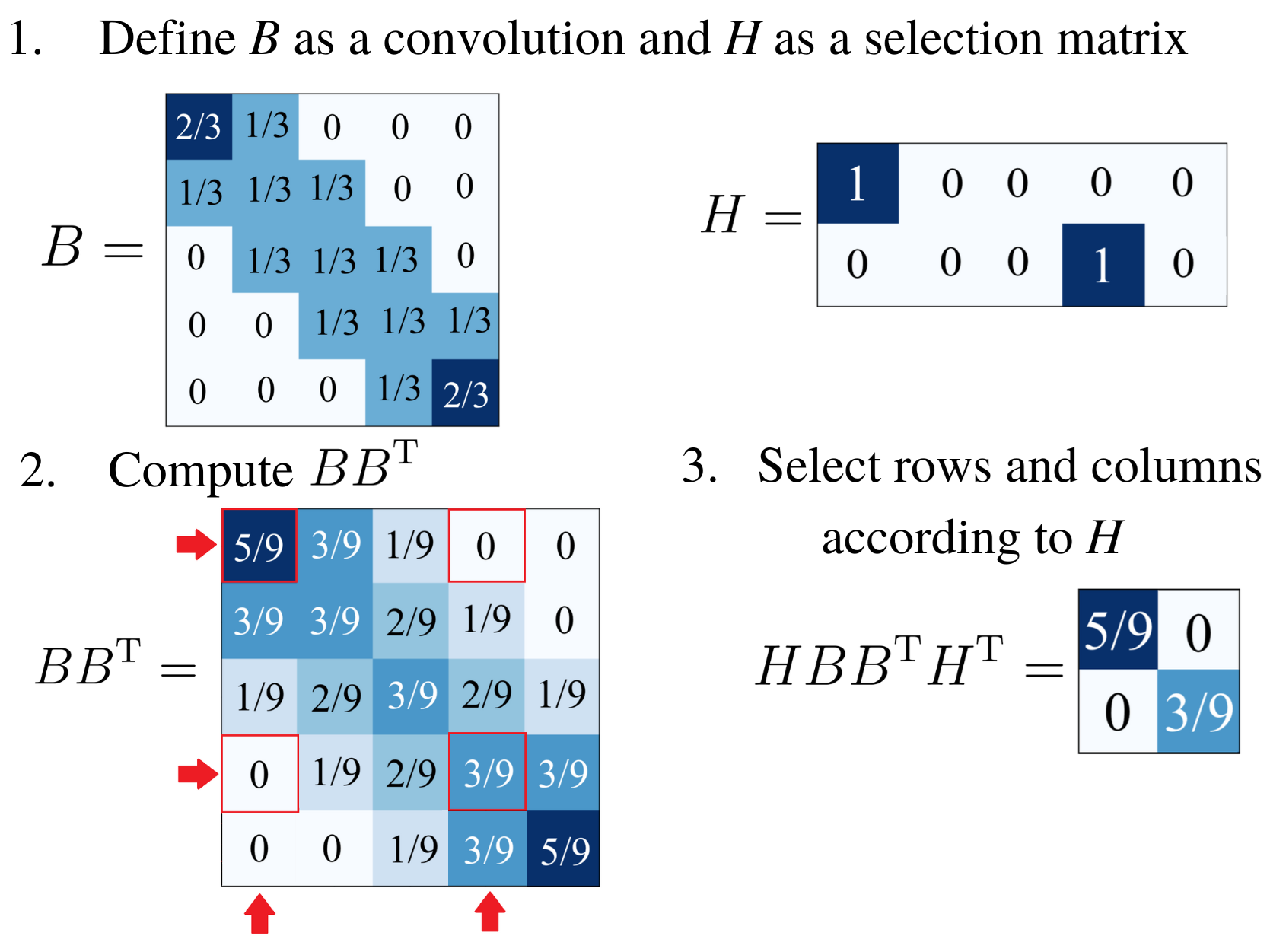}
  \caption{Example visualization demonstrating that $HBB^\text{T}H^\text{T}$ is a diagonal matrix based on our construction of $B$ and $H$. For simplicity, we show this result on a 1D example. Step 1 shows this example's choice of $B$ and $H$. The $B$ matrix above reflects a 1D convolutional kernel of length 3 with equal size weights with replication padding, and the choice of $H$ assumes the first and fourth element of the length 5 state vector is observed. Step 2 computes the matrix $BB^\text{T}$, which is now a symmetric matrix. The rows and columns with red arrows correspond to the indices that are observed based on our choice of $H$. As an aside, the matrix $BB^\text{T}H^\text{T}$ can be seen by retaining the columns with arrows shown in $BB^\text{T}$. Finally, Step 3 computes $HBB^\text{T}H^\text{T}$, which is a diagonal matrix. This matrix is constructed from retaining only the elements that have arrows in its row and column, as shown in the $BB^\text{T}$ matrix. We note that in the construction of $B$, centered at any particular index, $BB^T$ allows for a nonzero correlation of other  indices at most two spaces away. However, our choice of $H$ observes every third state, meaning that information from each observed index does not influence the state at another observation index since the two indices have 0 estimated covariance. Therefore, $HBB^\text{T}H^\text{T}$ can be stored as a vector of size 2 rather than a 2x2 matrix.} \label{fig:bbt}
\end{figure*}

In order to maximize computational efficiency, we carefully chose $C$ according to our gridded observations so that $(HCH^\text{T} + R)$ is a diagonal matrix, avoiding a matrix inverse that costs $\mathcal{O}(d_y^3)$ and instead computes the inverse of a vector of scalars, an operation with $\mathcal{O}(d_y)$ computational cost. Since $R$ is already assumed to be diagonal, we only need to construct $C$ to ensure that $HCH^\text{T}$ is diagonal. Intuitively, we want to construct $C$ such that any two distinct observation locations have 0 covariance. To ensure this property, we construct the background covariance $C$ based on $BB^\text{T}$, where $B$ is a convolution with kernel $W^{(k)}$, where $k=8,10,18,$ and 20 in our experiments. Constructed $C$ via $BB^\text{T}$ ensures that no matter our choice of $B$, the resulting $C$ is a proper covariance matrix. The convolution $B$ is applied to a state $X_t$ for any time $t$ and applies replication padding at the boundaries, which is naive to the fact that this data is collected over a sphere. Future implementations can implement a padding that reflects that the resulting image lies on a sphere.

Figure \ref{fig:bbt} visually justifies why our choice of $HCH^\text{T}$ is diagonal, allowing for us to avoid a computationally expensive $d_y$ matrix inversion. Figure \ref{fig:bbt} contains a 1D example showing that given a choice of $B$ and $H$ consistent with our experimental setting, $HCH^\text{T}$ is diagonal. Since we assume that our observations always lie on a regular grid, and given that our choice of $C$ guarantees that observation information will not propagate to another observation's location, $HCH^\text{T}$ is diagonal.

\section{Proof of Theorem \ref{thm:stability}}\label{app:proof}

 This appendix contains the proof of Theorem \ref{thm:stability}. The proof follows closely those of Theorem 3.2 in \cite{Moodey2013} and Theorem 9.2 in \cite{sanz2023inverse}. 
\begin{proof}
Throughout the proof, $c$ will denote a constant that may change from line-to-line. First, we introduce $x_t^o$ as the ``operational'' 3DVar analysis using observations $y_t$ and the ``true'' dynamics map $\mathcal{F}.$ Specifically, for $t=0$ we set $x_0^o = x_0^s,$ and then we recursively define
\begin{equation}
    x_t^o = (I-KH)\mathcal{F}(x_{t-1}^o) + Ky_t, \qquad t \ge 1.
\end{equation}

In operational weather settings, $\mathcal{F}$ is a NWP model.

The main idea of this proof is to decompose the error $\lVert x^s_t - x_t^\text{true}\rVert$ into two components:
\begin{enumerate}[label=($\alph*$)]

    \item filtering error with 3DVar using the true dynamics $\mathcal{F};$ and
    \item the distance between analyses from 3DVar using the true dynamics $\mathcal{F}$ and 3DVar using the surrogate dynamics $\mathcal{F}_s$.
\end{enumerate}

This decomposition can formally be written as 
\begin{equation}
    \lVert x_t^s - x_t^\text{true}\rVert \leq \underbrace{\lVert x_t^o - x_t^\text{true}\rVert}_{(a)} + \underbrace{\lVert x_t^s - x_t^o\rVert}_{(b)}. \label{eq:error_decomp}
\end{equation}

\noindent\textbf{($a$) Filtering error with 3DVar using the true dynamics $\mathcal{F}$.}

First focusing on term $(a)$ in inequality \eqref{eq:error_decomp}, we can rewrite $x_{t}^\text{true}$ and $x_t^o$ as
\begin{align}
    x_t^\text{true} & = (I-KH)\mathcal{F}(x_{t-1}^\text{true}) + KH \mathcal{F}(x_{t-1}^\text{true}) \label{eq:aux1} \\
    x_t^o & = (I-KH)\mathcal{F}(x_{t-1}^o) + KH\mathcal{F}(x_{t-1}^\text{true}) + \gamma K\eta_t, \label{eq:aux2}
\end{align}
where we use Assumption \ref{assumption:obs} to write $KH\mathcal{F}(x_{t-1}^o) = KH\mathcal{F}(x_{t-1}^\text{true}) +\gamma  K\eta_{t}$. Subtracting \eqref{eq:aux1} from \eqref{eq:aux2}, we obtain
\begin{align*}
    x_t^o - x_t^\text{true} & = (I-KH)\left[\mathcal{F}(x_{t-1}^o) - \mathcal{F}(x_{t-1}^\text{true})\right] + \gamma K\eta_t \\
    \ & = \left[\int_0^1 (I-KH)D\mathcal{F}(z x_{t-1}^o + (1-z)x_{t-1}^\text{true})dz\right](x_{t-1}^o - x_{t-1}^\text{true}) + \gamma K \eta_t,
\end{align*}
where we apply the mean-value-theorem for vector-valued functions to the first term. We then take the norm of both sides of the inequality
\begin{align*}
    \lVert x_t^o - x_t^\text{true}\rVert & \le \left[\int_0^1\lVert(I-KH)D\mathcal{F}(zx_{t-1}^o + (1-z)x_{t-1}^\text{true})\rVert dz\right]\lVert x_{t-1}^o - x_{t-1}^\text{true}\rVert + \gamma \lVert K\eta_t\rVert \\
    \ & \leq \lambda \lVert x_{t-1}^o - x_{t-1}^\text{true}\rVert + \gamma \lVert K\eta_t\rVert,
\end{align*}
where we use the assumption in inequality \eqref{assumption:bound_deriv} in Theorem \ref{thm:stability}. By further taking the expectation of both sides of the inequality, 
$$\mathbb{E}\lVert x_t^o - x_t^\text{true}\rVert \leq \lambda \mathbb{E}\lVert x_{t-1}^o - x_{t-1}^\text{true}\rVert + c \gamma,$$
where we assume that the scaled measurement noise $\lVert K\eta_t\rVert$ is bounded above by some constant $c>0.$ 
Therefore, we recursively deduce that
\begin{align*}
    \mathbb{E}\lVert x_t^o - x_t^\text{true}\rVert &
    \le \lambda \Bigl( \lambda \mathbb{E}\lVert x_{t-2}^o - x_{t-2}^\text{true} \rVert  + c \gamma \Bigr) + c \gamma \\
   &= \lambda^2 \mathbb{E}\lVert x_{t-2}^o - x_{t-2}^\text{true} \rVert +  c \gamma \lambda + c \gamma  \le \cdots \le  \lambda^t \mathbb{E}\lVert x_{0}^o - x_{0}^\text{true} \rVert + c \gamma \sum_{i=0}^{t-1} \lambda^i .
\end{align*}
Finally, since $\lambda \in (0,1)$, we conclude that
\begin{equation}
    \lim_{t\to\infty} \sup \mathbb{E}\lVert x_t^o - x_t^\text{true}\rVert \leq \frac{\gamma c}{1-\lambda}.\label{ineq:part_a}
\end{equation}

\noindent\textbf{$(b)$ Distance between analyses from 3DVar using the true dynamics $\mathcal{F}$ and analyses from 3DVar using the surrogate dynamics $\mathcal{F}_s$.}

Now, shifting focus to finding an upper-bound of term $(b)$ in inequality \eqref{eq:error_decomp}, we can similarly rewrite $x_t^s$ and $x_t^o$ as 
\begin{align*}
    x_t^s & = (I-KH)\mathcal{F}_s(x_{t-1}^s) + KH\mathcal{F}_s(x_{t-1}^\text{true}) + \gamma K\eta_t, \\
    x_t^o & = (I-KH)\mathcal{F}(x_{t-1}^o) + KH\mathcal{F}(x_{t-1}^\text{true}) + \gamma K\eta_t,
\end{align*}
where we again use Assumption \ref{assumption:obs}. By writing out the expression for the distance between the analyses from 3DVar using the surrogate dynamics $\mathcal{F}_s$ and from 3DVar using the true dynamics $\mathcal{F}$, we obtain the following expression
\begin{align*}
    x_t^s-x_t^o & = (I-KH)\left[\mathcal{F}_s(x_{t-1}^s) - \mathcal{F}(x_{t-1}^o)\right] \\
    \ & = (I-KH)\left[\mathcal{F}_s(x_{t-1}^s)  - \mathcal{F}(x_{t-1}^s) \right]+ (I-KH)\left[\mathcal{F}(x_{t-1}^s)- \mathcal{F}(x_{t-1}^o)\right]. 
\end{align*}
 Utilizing the assumption in \eqref{assumption:bound_surrogate} to upper-bound the first term and the mean-value-theorem for vector-valued functions and the assumption of inequality \eqref{assumption:bound_deriv} in Theorem \ref{thm:stability} to upper-bound the second term, we obtain that
\begin{align*}
    \lVert x_{t}^s - x_{t}^o\rVert & \leq \varepsilon + \left[\int_0^1\lVert(I-KH)D\mathcal{F}(z x_{t-1}^s + (1-z)x_{t-1}^o)\rVert dz\right]\lVert x_{t-1}^s - x_{t-1}^o\rVert \\
    \ & \leq \varepsilon + \lambda \lVert x_{t-1}^s - x_{t-1}^o\rVert,
\end{align*}
By taking the expectation of both sides of the inequality, we obtain the inequality
$$\mathbb{E}\lVert x_{t}^s - x_{t}^o \rVert\leq \varepsilon + \lambda \mathbb{E}\lVert x_{t-1}^s - x_{t-1}^o\rVert.$$
Therefore,  by the same recursive argument used to bound term $(a)$, we deduce that
\begin{equation}
    \lim_{t\to\infty}\sup\mathbb{E}\lVert x_{t}^s - x_{t}^o \rVert \leq \frac{\varepsilon}{1-\lambda}.\label{ineq:part_b}
\end{equation}

\noindent\textbf{Combining the upper-bounds for $(a)$ and $(b)$ in inequality \eqref{eq:error_decomp}.}

By combining the two upper-bounds from \eqref{ineq:part_a} and \eqref{ineq:part_b}, we have the final result 
\begin{equation*}
    \lim_{t\to\infty}\sup \mathbb{E}\lVert x_{t-1}^s - x_t^\text{true}\rVert \leq c\left(\frac{\gamma + \varepsilon}{1-\lambda}\right).
\end{equation*}
\end{proof}

\section{Evaluation metrics}\label{app:metrics}
    
We evaluate our results using three metrics: latitude-weighted root mean square error (RMSE), latitude-weighted anomaly correlation coefficient (ACC), and continuous ranked probability score (CRPS). We formulate RMSE and ACC based on the definitions provided in \cite{Rasp2020}. 

The latitude-weighted RMSE at time $t$ across $f$ different features is defined in equation \eqref{eq:rmse} as 

\begin{equation}
    \text{RMSE}_t = \frac{1}{f}\sum_{f'=1}^f\sqrt{\frac{1}{N_\text{lat}N_\text{lon}}\sum_{i=1}^{N_\text{lon}}\sum_{j=1}^{N_\text{lat}} L(j)(\{x_t^s\}_{f',i,j} - \{x^\text{true}_t\}_{f',i,j})^2},
    \label{eq:rmse}
\end{equation}

\noindent where $N_\text{lat}$ is the number of latitudes, $N_\text{lon}$ is the number of longitudes, and $L(j)$ is the latitude weighting factor for the $j$th latitude index, defined in equation \eqref{eq:latitude_weighting} as 

\begin{equation}
    L(j) = \frac{\text{cos}(\text{lat}(j))}{\frac{1}{N_\text{lat}}\sum_{j=1}^{N_\text{lat}}\text{cos}(\text{lat}(j))}.
    \label{eq:latitude_weighting}
\end{equation}

We additionally utilize the latitude-weighted ACC at time $t$
across $f$ different features, defined in equation \eqref{eq:acc} as 

\begin{equation}
    \text{ACC}_t = \frac{1}{f}\sum_{f'=1}^f \frac{\sum_{i=1}^{N_\text{lon}}\sum_{j=1}^{N_\text{lat}}L(j) \{x_t^s\}_{f',i,j} \{x_t^\text{true}\}_{f',i,j} }{ \sqrt{\sum_{i=1}^{N_\text{lon}} \sum_{j=1}^{N_\text{lat}} (\{x_t^s\}_{f',i,j})^2  \sum_{i=1}^{N_\text{lon}} \sum_{j=1}^{N_\text{lat}} L(j)(\{x_t^\text{true}\}_{f',i,j})^2 } }
    \label{eq:acc}
\end{equation}

In both equations \eqref{eq:rmse} and \eqref{eq:acc}, $\{x_t^s\}_{f',i,j}$ is defined as an estimate, either through forecasting or filtering, of $\{x_t^\text{true}\}_{f',i,j}$ for feature $f'$ at latitude $j$ and longitude $j$, and $\{x_t^\text{true}\}_{f',i,j}$ is the ground truth ERA5 data for feature $f'$ at latitude $j$ and longitude $i$. In all of our experiments, $N_\text{lat}=720$ and $N_\text{lon}=1440$.

We compute the CRPS \citep{Matheson1976}, which is defined as follows, 

\begin{equation}
    \text{CRPS}(F, y) = \int_{-\infty}^{\infty} \big(F(y) - \mathds{1}(y-x)\big)^2 dx,
\end{equation}
\noindent where $y$ is an observation, $F$ is the CDF of the forecast distribution, and $\mathds{1}$ is the Heavside function. In our work, the $F$ we provide is an empirical CDF of the forecast distribution based on an ensemble.

\section{Assimilation visualizations for all atmospheric features}\label{sec:assimilation_vis}

\begin{figure*}
  \centering
  \includegraphics[width=13cm]{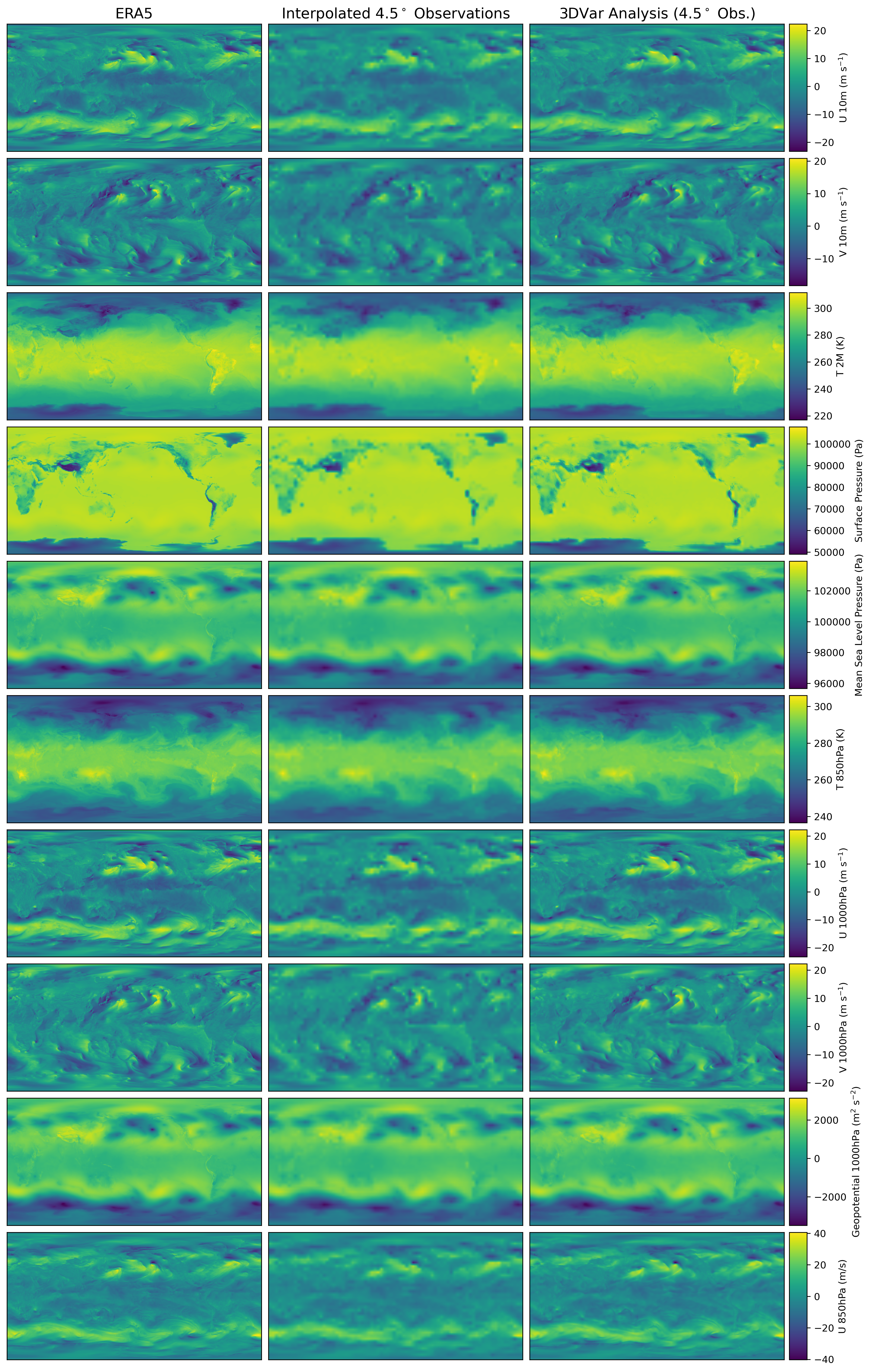}
  \caption{Visualization of the ground truth ERA5 data, interpolated $4.5^\circ$ ERA5 observations with standardized $N(0,0.0001 I_{d_y})$ distributed additive errors, and our 3DVar analysis using this observational data and FourCastNet for 10 different atmospheric features at the end of our assimilation horizon, December 31, 2023 at 18:00 UTC.} \label{fig:assim_all_features_part1}
\end{figure*}

\begin{figure*}
  \centering
  \includegraphics[width=13cm]{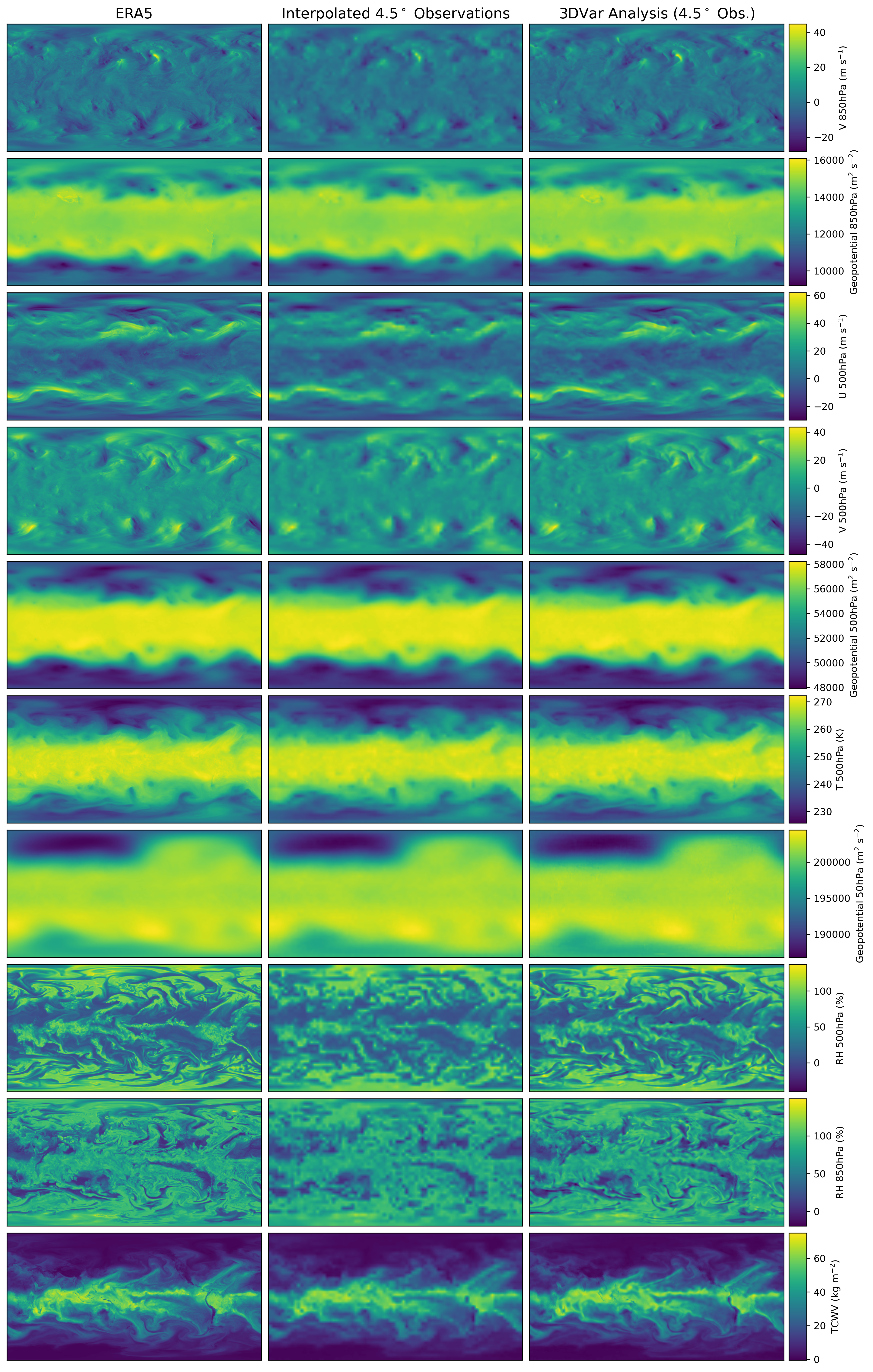}
  \caption{Visualization of the ground truth ERA5 data, interpolated $4.5^\circ$ ERA5 observations with standardized $N(0,0.0001 I_{d_y})$ distributed additive errors, and our 3DVar analysis using this observational data and FourCastNet for 10 different atmospheric features at the end of our assimilation horizon, December 31, 2023 at 18:00 UTC.} \label{fig:assim_all_features_part2}
\end{figure*}

Figures \ref{fig:assim_all_features_part1} and \ref{fig:assim_all_features_part2} visualize the ground truth ERA5 data, interpolated $4.5^\circ$ noisy observations, and 3DVar analyses using these $4.5^\circ$ observations for all 20 atmospheric features. The 3DVar analyses were constructed from 365 days of assimilating sparse, noisy $4.5^\circ$ resolution observations every 6 hours. 

The rows of Figure \ref{fig:assim_all_features_part1} visualize, in order, U-component wind speed at 10m above the surface, V-component wind speed at 10m above the surface, temperature at 2 meters above the surface, surface pressure, mean sea level pressure, temperature at pressure level 850 hPa, U-component wind speed at pressure level 1000 hPa, V-component wind speed at pressure level 1000 hPa, geopotential at pressure level 1000 hPa, and U-component wind speed at pressure level 850 hPa.

The rows of Figure \ref{fig:assim_all_features_part2} visualize, in order, V-component wind speed at pressure level 850 hPa, geopotential at pressure level 850 hPa, U-component wind speed at pressure level 500 hPa, V-component wind speed at pressure level 500 hPa, geopotential at pressure level 500 hPa, temperature at pressure level 500 hPa, geopotential at pressure level 50 hPa, relative humidity at at pressure level 500 hPa, relative humidity at pressure level 850 hPa, and total column water vapor.

\section{Sample 48 hour ensemble forecasts for Typhoon Mawar, 2023} \label{app:ens_mawar}

\begin{figure*}
  \centering
  \includegraphics[width=15cm]{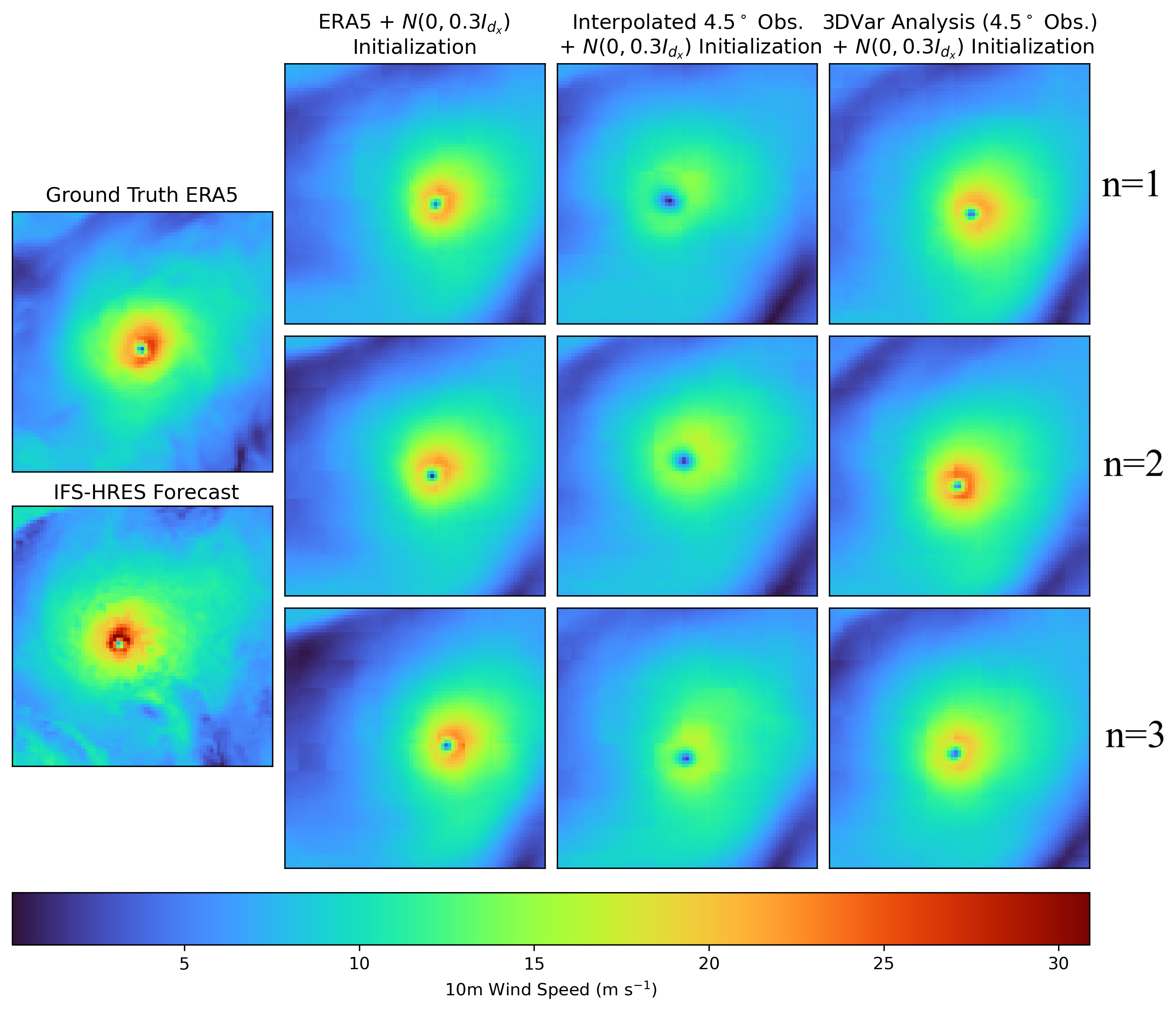}
  \caption{Visualization of three ensemble members' 48 hour forecasts of 10m wind speed for May 25, 2023 using three different initializations: noisy ground truth ERA5 data, noisy interpolated $4.5^\circ$ observations, and a 3DVar analysis ($4.5^\circ$ observations). For visual reference, we include ground truth ERA5 and the IFS-HRES forecast of 10m wind speed on May 25, 2023.} \label{fig:mawar_ws_ens}
\end{figure*}

\begin{figure*}
  \centering
  \includegraphics[width=15cm]{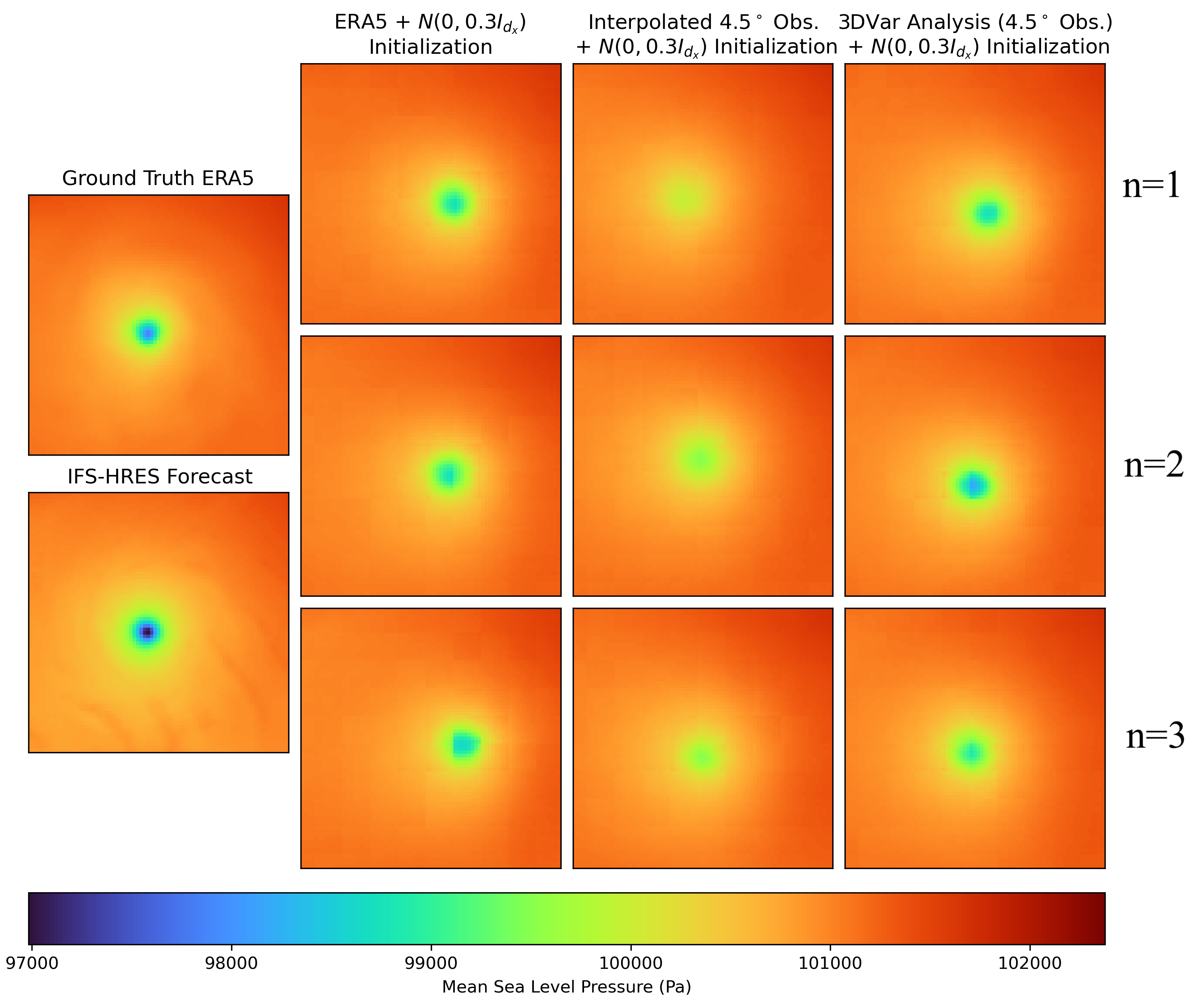}
  \caption{Visualization of three ensemble members' 48 hour forecasts of mean sea level pressure for May 25, 2023 using three different initializations: noisy ground truth ERA5 data, noisy interpolated $4.5^\circ$ observations, and a 3DVar analysis ($4.5^\circ$ observations). For visual reference, we include ground truth ERA5 and the IFS-HRES forecast of 10m wind speed on May 25, 2023.} \label{fig:mawar_pressure_ens}
\end{figure*}

To supplement the single ensemble member visualizations in Figures \ref{fig:mawar_ws_one_ens} and \ref{fig:mawar_pressure_one_ens}, we include visualizations for the 48 hour forecasts of three other randomly selected ensemble members for each of our initial conditions in Figures \ref{fig:mawar_ws_ens} and \ref{fig:mawar_pressure_ens}. As mentioned in the main text, the forecasts from the interpolated $4.5^\circ$ observations generally tend to result in less extreme predictions compared to the ground truth or the forecasts from initializing with noisy ground truth ERA5 data or our 3DVar analysis (4.5$^\circ$ observations).

\end{document}